  \documentstyle[12pt,epsfig]{article}

\def\journal#1, #2, #3, #4 { {\sl #1~}{\bf #2~} (#3)  #4 }

\def\prd{\journal Phys. Rev. D, }

\def\prl{\journal Phys. Rev. Lett., }

\def\cmp{\journal Comm. Math. Phys., }

\def\np{\journal Nucl. Phys., }

\def\pl{\journal Phys. Lett., }

\def\annp{\journal Ann. Phys. (N.Y.), }

\def\ijmp{\journal Int. J. Mod. Phys., }

\catcode`\@=11
\def\marginnote#1{}
\newcount\hour
\newcount\minute
\newtoks\amorpm
\hour=\time\divide\hour by60
\minute=\time{\multiply\hour by60 \global\advance\minute
by-\hour}\edef\standardtime{{\ifnum\hour<12
\global\amorpm={am}%
        \else\global\amorpm={pm}\advance\hour by-12 \fi
        \ifnum\hour=0 \hour=12 \fi
        \number\hour:\ifnum\minute<10
0\fi\number\minute\the\amorpm}}
\edef\militarytime{\number\hour:\ifnum\minute<10
0\fi\number\minute}

\def\draftlabel#1{{\@bsphack\if@filesw {\let\thepage\relax
   \xdef\@gtempa{\write\@auxout{\string
      \newlabel{#1}{{\@currentlabel}{\thepage}}}}
Ebenen
}\@gtempa
   \if@nobreak \ifvmode\nobreak\fi\fi\fi\@esphack}
        \gdef\@eqnlabel{#1}}
\def\@eqnlabel{}
\def\
Ebenen
@vacuum{}
\def\draftmarginnote#1{\marginpar{\raggedright\scriptsize\tt#1}}
\def\draft{\oddsidemargin -.5truein
        \def\@oddfoot{\sl preliminary draft \hfil
        \rm\thepage\hfil\sl\today\quad\militarytime}
        \let\@evenfoot\@oddfoot \overfullrule 3pt
        \let\label=\draftlabel
        \let\marginnote=\draftmarginnote

\def\@eqnnum{(\theequation)\rlap{\kern\marginparsep\tt\@eqnlabel}%
\global\let\@eqnlabel\@vacuum}  }

\def\numberbysection{\@addtoreset{equation}{section}
        \def\theequation{\thesection.\arabic{equation}}}

\def\underline#1{\relax\ifmmode\@@underline#1\else
 $\@@underline{\hbox{#1}}$\relax\fi}

\catcode`@=12
\relax
\numberbysection
\def\beq{\begin{equation}}
\def\eeq{\end{equation}}
\def\beqa{\begin{eqnarray}}
\def\eeqa{\end{eqnarray}}
 \def\nnn{\nonumber \\}

\def\hhat{{\widehat h}}
\def\lfloorhat{{\hat \lfloor}}
\def\rfloorhat{{\hat \rfloor}}

\def\Jhat{{\widehat J}}
\def\Khat{{\widehat K}}
\def\Je{J^e{}}

\def\Jehat{{\Jhat^e}{}}

\def\kappahat{{\widehat \kappa}}

\def\mhat{{\widehat m}}
\def\nhat{{\widehat n}}

\def\Vhat{{\widehat V}}
\def\Uhat{{\widehat U}}

\def\xhat{{\widehat x}}

\def\Vhat{{\widehat V}}

\def\muhat{{\widehat \mu}}

\def\qhat{{\widehat q}}

\def\varpihat{{\widehat \varpi}}

\def\varpib{{\overline \varpi}}
\def\varthetab{{\overline \vartheta}}
\def\kappab{{\overline \kappa}}

\def\mhat{{\widehat m}}
\def\nhat{{\widehat n}}
\def\phat{{\widehat p}}

\def\Shat{{\widehat S}}

\def\xb{{\bar x}}
\def\Jb{{\bar J}}

\def\Sb{{\overline  S}}
\def\zb{{\bar z}}

\def\mb{{\bar m}}

\def\Shatb{{\widehat {\overline S}}}
\def\Ub{{\overline U}}

\def\nub{\bar \nu}

\def\Jgen#1 {  {\underline J_{#1}} }
\def\Kgen#1 {  {\underline K_{#1}} }
\def\Jgenp#1 #2 {(J_{#1}+{#2},\Jhat_{#1})}
\def\Jgenm#1 #2 {(J_{#1}-{#2},\Jhat_{#1})}
\def\Jg#1 {J_{#1},\Jhat_{#1}}
\def\Jgp#1 #2 {J_{#1}+{#2},\Jhat_{#1}}
\def\Mgen#1 {{\underline M_{#1}}}
\def\mgen#1 {{\underline m_{#1}}}
\def\xgen{{\underline x}}
\def\ms{m^\circ{}}
\def\msone{m_1^\circ}
\def\mstwo{m_2^\circ}
\def\mshat{ {\mhat^\circ}{}}
\def\msonehat{{\mhat_1^\circ}{}}
\def\mstwohat{{\mhat_2^\circ}{}}
\def\mssum{m_{12}^\circ{}}
\def\mssumhat{{\mhat_{12}^\circ}{}}
\def\Vt{{\widetilde V}}
\def\Vthat{\widehat{\widetilde V}}
\def\Vtb{\overline{\Vt}}

\def\fin{\end{document}}

\def\Jge{{\underline J}}
\def\mge{{\underline m}}

\def\Jgen#1 {  {\underline J_{#1}} }
\def\Jgenp#1 #2 {(J_{#1}+{#2},\Jhat_{#1})}
\def\Jgenm#1 #2 {(J_{#1}-{#2},\Jhat_{#1})}
\def\Jg#1 {J_{#1},\Jhat_{#1}}
\def\Jgp#1 #2 {J_{#1}+{#2},\Jhat_{#1}}
\def\Mgen#1 {{\underline M_{#1}}}
\def\me{m^e{}}

\def\fusV#1,#2,#3,#4,#5,#6 {f_V(
\Jgen{#1} ,
\Jgen{#2} ,
\Jgen{#3} ,
\Jgen{#4} ,
\Jgen{#5} ,
\Jgen{#6} )}

\def\sixjxi#1,#2,#3,#4,#5,#6
{{\left\{\left . \!\! \,^{#1}_{#2}\,^{#3}_{#4}
\right | \!\, ^{#5}_{#6}\right\}}}

\def\gaghat{{\hat {\bigl \{}}}
\def\gadhat{{\hat {\bigr \}}}}
\def\Gaghat{{\hat {\Bigl \{}}}
\def\Gadhat{{\hat {\Bigr \}}}}

\def\bverthat{{\hat {\Bigl \vert}}}

\begin{document}
\tolerance 2000
\hbadness 2000
\begin{titlepage}

\nopagebreak \begin{flushright}

LPTENS--93/40 \\
hep-th/9405136
 \\
    May 1994
\end{flushright}

\vglue 3  true cm
\begin{center}
{\large \bf
CONTINUOUS SPINS IN 2D GRAVITY:  \\
\medskip
CHIRAL VERTEX OPERATORS AND LOCAL FIELDS}\\
\vglue 1.5 true cm
{\bf Jean-Loup~GERVAIS}\\
\medskip
and \\
\medskip
{\bf Jens SCHNITTGER}{\footnotesize\footnote{supported by DFG}}\\
\medskip
{\footnotesize Laboratoire de Physique Th\'eorique de
l'\'Ecole Normale Sup\'erieure\footnote{Unit\'e Propre du
Centre National de la Recherche Scientifique,
associ\'ee \`a l'\'Ecole Normale Sup\'erieure et \`a
l'Universit\'e
de Paris-Sud.},\\
24 rue Lhomond, 75231 Paris CEDEX 05, ~France}.
\end{center}
\vfill
\vglue 1 true cm
\begin{abstract}
\baselineskip .4 true cm
{\footnotesize
\noindent
We construct  the
exponentials of the Liouville
field with continuous powers
within the operator approach. Their chiral decomposition is realized
using the explicit Coulomb-gas operators we introduced earlier.
{}From the quantum-group viewpoint,
they are related to semi-infinite highest or lowest weight representations
with continuous spins.
The Liouville field itself is defined, and the canonical
commutation relations verified, as well as the validity
 of  the quantum
Liouville field equations. In a second part, both screening
charges are considered.
The braiding of the chiral components
 is derived and shown to agree with the ansatz  of a
parallel paper of  J.-L. G. and Roussel: for continuous spins the
quantum group structure $U_q(sl(2)) \odot U_{\qhat}(sl(2))$ is a
non trivial extension of $U_q(sl(2))$ and $U_{\qhat}(sl(2))$.
We construct the corresponding generalized exponentials and the generalized
Liouville field.
}
\end{abstract}
\vfill
\end{titlepage}
\section{Introduction}
Until recent times, the  progress in  understanding
the structure of two-dimensional gravity from the operator  point
of view\cite{B}-\cite{CGR2}, was based on the detailed study of the
monodromy properties of the Virasoro
null-vector equations, whose link with the quantum group
$U_q(sl(2))$ is completely understood by now\cite{CGR1}\cite{CGR2}.
The degenerate fields  correspond to
standard  representations of this quantum group, \footnote{
 or  of its
suitable extension when the two types
of screening charges are included,}
with positive half-integer spins\cite{G1}.
The quantum group structure of the theory
turns out to govern not only
the chiral operator
algebra\cite{B}\cite{G1}\cite{G3}\cite{CGR1}\cite{CGR2},
but also the reconstruction of ( the exponential
of) the Liouville  field\cite{G5} $\exp (-J\alpha_- \Phi)$,
which is  simply  the $U_q(sl(2))$-singlet made out of two
representations of spin $J$. It also opens the way
towards understanding 2D gravity in the
strong coupling regime\cite{G3}.
However, the study of representations with half-integer spins
 does not
by far  answer all the physical questions we want to ask
about 2D gravity.
In particular, modular invariance in the strong coupling regime
forces us to consider operators with
quantum-group spins which are rational,
but  not halves of integers\cite{GR}, and the possibility of
defining the Liouville field itself, and not just some of its exponentials,
is realized only if we can define the Liouville exponential
with continuous $J$, so that we may let $\Phi =-{d\over dJ}|_{J
=0} \exp (-J\alpha_- \Phi)/ \alpha_- $. The basic difficulty in
going away from half-integer spins is that one no longer deals with
degenerate fields satisfying null-vector equations. In a
recent letter we have shown\cite{GS1}
 how to solve this problem, at least
concerning the braiding, by using an operator Coulomb-gas realization,
where  the braiding matrix turned out to be computable
from a simple quantum
mechanical problem, which we could solve in closed form.
The point of the present article is to go further in the same direction.
Another line of attack has recently been followed in ref.\cite{GR},
where the fusing and braiding matrices are generalized using the
scheme\cite{MS} of Moore and Seiberg, and by
requiring that the polynomial equations still hold for non-integer
$2J$'s. As we will see the two methods agree.

This article is organized as follows. In section 2 we recall some
background material. Section 3 is devoted to the case of a single
screening charge, where the quantum group structure is the
standard $U_q(sl(2))$. Besides completing the discussion
of ref.\cite{GS1}
for the chiral algebra, we construct Liouville
exponentials for arbitrary $J$ and give an expression for the Liouville
field itself. It is shown that the canonical equal-time commutation
relations  as well as the quantum equations of motion are satisfied.
 We discuss the periodicity properties of our definition
 of the Liouville exponentials
resp. of the Liouville field and their connection with the presence
of singularities in the elliptic sector. The preservation on the
quantum level
of the symmetry under the exchange of the two equivalent Backlund free
fields\cite{GN3} is derived for half-integer spins.
In section 4, we consider both screening charges
together. The corresponding quantum group structure was noted
$U_q(sl(2))\odot U_{\qhat}(sl(2))$ in ref.\cite{G3}. In the degenerate
case, the primary fields of   spins $J$ and $\Jhat$ are of the type
$(2\Jhat+1, 2J+1)$ in the BPZ classification. Then  the braiding of a
$(1, 2J+1)$ field with a $(2\Jhat+1, 1)$ field is a simple phase, so
that this $\odot$ symbol represents a   sort of graded tensor-product.
The situation is completely different for
 continuous spins, and this comes out very
naturally from our approach where it is shown that, in this general
situation, the relevant parameter is the quantity called effective
spin given by $J^e=J+\Jhat \pi/h$ ($h$ is defined by $q=\exp (ih)$ as
usual). Whereas, for half integer $J$ $\Jhat$ and generic $h$,
it is equivalent to specify $J$ $\Jhat$ or $J^e$, the former loose
their meaning  for continuous spins.
Thus the $U_q(sl(2))\odot U_{\qhat}(sl(2))$
structure has novel features which we investigate in detail.
In the  parallel study of this general situation\cite{GR},
extending   the
polynomial equations  of the Moore Seiberg approach has allowed
to write educated guesses for the fusion and braiding matrices
that are remarkably simple in terms of these effective spins.
Although it is unable, at present to directly
deal with fusion, our approach
allows us  to actually derive the braiding matrices, which completely
agree with the expression of ref.\cite{GR}. Assuming the standard
relationship  between braiding and fusion matrices, this
fully determines the operator algebra of the chiral primary fields
for continuous spins, and  agrees with ref.\cite{GR}. It is
also used to construct the Liouville exponential in this  most
general
situation.

\section{Trying to be self-contained}
\label{self-contained}
Let us  rederive  some background material about
Liouville theory, in order to introduce  the coming discussion.
The solutions of the classical Liouville dynamics, which is
described by the action
\begin{equation}
S={1\over 8\pi}\int d\tau d\sigma  \{ (\partial_\tau \Phi )^2
-(\partial_\sigma \Phi )^2 -\mu^2 e^{2 \sqrt{\gamma} \Phi} \},
\label{2.1}
\end{equation}
are given by \hfill
\begin{equation}
2\sqrt{\gamma} \Phi =
\ln \left [{8\over \mu^2\sqrt{\gamma}}
{A'(u)B'(v)\over  (A(u)-B(v))^2 }\right ]
\qquad , \qquad u=\tau +\sigma, \quad v=\tau -\sigma
\label{2.2}
\end{equation}
with $A$ and $B$ arbitrary functions, and $\sigma \in [0,2\pi ]$.
The coupling constant is noted $\gamma$.
We have redefined $\Phi \to 2\sqrt{\gamma}\Phi$ in order to agree with
the classical limit of
standard quantum normalizations, where $2\sqrt{\gamma}$ is the limit of the
screening charge $\alpha_-$.
Eq.\ref{2.2} is invariant under the projective transformations
\begin{equation}
A\to {aA+b\over cA+d} \qquad B\to {aB+b\over cB+d} \qquad ,
\label{2.3}
\end{equation}
$a,b,c,d$ complex, which on the quantum level gives rise to the
$U_q(sl(2))$ quantum group symmetry. Introducing the chiral
fields\footnote{In this formula, contrary to the rest of the
article, we of course  use ordinary binomial coefficients.}
\begin{eqnarray}
f_m^{(J)}&=& \sqrt{\scriptstyle {2J \choose J+m}}
(A'^{-1/2})^{J-m} (AA'^{-1/2})^{J+m},\nnn
\bar f_m^{(J)}&=&\sqrt{\scriptstyle {2J \choose J+m}}
(B'^{-1/2})^{J+m} (BB'^{-1/2})^{J-m},
\label{2.4}
\end{eqnarray}
we can write the Liouville exponentials as
\begin{equation}
e^{-2J \sqrt{\gamma}\Phi} =\left ({\mu^2 \sqrt{\gamma}\over 8}\right )^J
\sum_{m=-J}^{J}  (-1)^{J+m}
f_m^{(J)}\bar f_m^{(J)}
\label{2.5}
\end{equation}
for any positive half-integer $J$. It is easy to verify that,
when $A$ and $B$ undergo  the M\"obius transformation
Eq.\ref{2.3},   the functions
$f_m^{(J)},\bar f_m^{(J)}$ transform as standard
 (finite-dimensional)
spin $J$ representations of $sl(2,\bf{C})$, and Eq.\ref{2.5} is the
corresponding singlet.
For general  $J$, Eq.\ref{2.5}
is still valid formally, with the $m-$sum extending to $+\infty$
or  to $-\infty$, depending on whether we work with the
semi-infinite  representations with $J+m$ positive integer,
or $J-m$ positive integer. In practice, Eq.\ref{2.5} is
an expansion  in $(A/B)^m$, and
highest-weight representations, with  $-\infty\leq m\leq J$
(resp. lowest-weight representations, with $-J \leq m\leq \infty$)  will
give a convergent expansion for $|A|> |B|$ (resp. $|B| > |A|$).
Both choices should
 represent the same function, since
they are just related by the particular $sl(2)$-transformation
\begin{equation}
A\to -1/A  \qquad B\to -1/B \quad ,
\label{2.6}
\end{equation}
which sends $m \to -m$ in $f_m^{(J)}, \bar f_m^{(J)}$.\footnote{Actually
the situation is somewhat more subtle for singular solutions,
as we will see later.}  Of course,
in the case of positive half-integer $J$, this amounts only to a
trivial permutation
of terms in the sum Eq.\ref{2.5}. For continous $J$, however, the highest
resp. lowest weight representations are representations only of the algebra
but not of the group, due to the multivaluedness of the $f_m^{(J)}$ under the
group operations Eq.\ref{2.3}. Consequently, the
transformation Eq.\ref{2.6} exchanges highest and lowest
weight representations. From the general point of view of Toda theory,
it can be regarded as representing the Weyl group symmetry\cite{ANPS}.

Periodicity of $\Phi$ implies that $A$ and $B$ must be periodic up to
a projective transformation, which is called the monodromy matrix.
In the elliptic and hyperbolic sectors of the theory, we can always
pick a representative of the equivalence class defined by Eq.\ref{2.3}
such that the monodromy matrix is diagonal, i.e. such that $A$ and
$B$ are periodic up to a multiplicative constant. In fact, there are
precisely two such representatives, related by Eq.\ref{2.6} which can
thus be viewed in this context as a kind of residual symmetry.
It is then possible to define two equivalent sets of chiral free
fields by\footnote{Compared to ref.\cite{GS2}, we have changed the notation
by replacing $\phi_1 \to \vartheta_1,  \  \phi_2 \to \vartheta_2,
\ -\bar\phi_1 \to \bar\vartheta_2,
\ -\bar\phi_2 \to \bar\vartheta_1$
.}
\begin{eqnarray}
\sqrt{{\gamma }}\vartheta_1(u):=
\ln A'^{-1/2}(u),&  \quad  \sqrt{{\gamma }}
\bar\vartheta_1(v):=
\ln BB'^{-1/2}(v), \nnn
\sqrt{{\gamma}}\vartheta_2(u):=\ln AA'^{-1/2}(u), &
\quad \sqrt{{\gamma}} \bar \vartheta_2(v):=
\ln B'^{-1/2}(v)
\label{freefields}
\end{eqnarray}
Indeed, one may show that the canonical Poisson brackets
 of Liouville theory
give the following free-field Poisson bracket relations
\begin{equation}
\Bigl \{\vartheta'_1(\sigma_1),\vartheta'_1(\sigma_2)
\Bigr \}_{\hbox {\footnotesize P.B.}}=
\Bigl \{\vartheta'_2(\sigma_1),\vartheta'_2(\sigma_2)
\Bigr \}_{\hbox {\footnotesize P.B.}}
=2\pi \,  \delta'(\sigma_1-\sigma_2)
\label{2.8}
\end{equation}
with similar relations for the bar components.
The explicit mode expansions in terms of zero modes and oscillators
are given by
\begin{eqnarray}
\phantom{1234567890}&\vartheta_j(u)=q^{(j)}_0+ p^{(j)}_0 u+
i\sum_{n\not= 0}e^{-in u}\, p_n^{(j)}\bigl / n,
\quad \phantom{j=1,\> 2}  \phantom{1234567890}\cr
\phantom{1234567890}&\bar\vartheta_j(v)=\bar q^{(j)}_0+\bar p_0^{(j)}v+
i\sum_{n\not= 0}e^{-in v}\, \bar p_n^{(j)}\bigl / n,\quad j=1,\> 2
\phantom{1234567890}\cr
\label{1.8}
\end{eqnarray}
{}From Eq.\ref{1.8} we see that
the periodicity properties of the $A$ and $B$ fields can
be parametrized by the zero mode momenta e.g. of $\vartheta_1$:
\beq
A(u+2\pi)=e^{-4\pi p_0^{(1)}\sqrt{\gamma}} A(u), \quad
B(v+2\pi)=e^{+4\pi \bar p_0^{(1)}\sqrt{\gamma}} B(v)
\label{2.9}
\eeq
The complete symmetry of the treatment of the theory
 under the exchange of
$\vartheta_1,\bar \vartheta_1$
and
$\vartheta_2,\bar \vartheta_2$ even on the quantum level  is  the
hallmark of Gervais-Neveu
quantization and guarantees the  preservation
of the residual  symmetry Eq.\ref{2.6}. From
Eqs.\ref{2.4}, \ref{freefields} we have that the fields $f_m^{(J)}$
can be written as products of exponentials of the $\vartheta_1$
and $\vartheta_2$ fields:
\beq
f_m^{(J)}=e^{(J-m)\sqrt{\gamma}\vartheta_1}e^{(J+m)\sqrt{\gamma}\vartheta_2}
\label{exprep}
\eeq
Though this form is, in principle,
accessible directly to quantization  for any $J$
and $m$ (cf. ref.\cite{GS1} ),  for the purposes of the present paper
it is more appropriate to work with an alternative Coulomb-gas-type
representation
in terms of one free field only. Let us consider the special cases
$m=\pm J$ of Eq.\ref{exprep} where we have
\beq
f^{(J)}_{-J}=({1\over \sqrt{A'}})^{2J}=
e^{ 2J\sqrt{\gamma}\vartheta_1}, \qquad
f^{(J)}_{J}=({A\over \sqrt{A'}})^{2J}=e^{ 2J\sqrt{\gamma}\vartheta_2},
\label{2.10}
\eeq
Using the  periodicity requirement Eq.\ref{2.9},
one easily derives the relations
\beqa
A(u)&=&
\left \{ e^{-4\pi p_0^{(1)} \sqrt{\gamma} }
\int_0^uf^{(-1)}_{1}(\rho)d\rho +
\int_u^{2\pi}f^{(-1)}_{1}(\rho)d\rho \right \} \left /
\left ( e^{-4\pi p_0^{(1)} \sqrt{\gamma} } -1\right )\right. \quad
\label{2.11} \\
{-1\over A(u)}&=&
\left \{ e^{-4\pi p_0^{(2)} \sqrt{\gamma} }
\int_0^uf^{(-1)}_{-1}(\rho)d\rho +
\int_u^{2\pi}f^{(-1)}_{-1}(\rho)d\rho \right \} \left  /
\left ( e^{-4\pi p_0^{(2)} \sqrt{\gamma} } -1 \right)\right.  \quad
\label{2.12}
\eeqa
Then we may rewrite Eq.\ref{2.4} as
\beq
f_m^{(J)}= \sqrt{\scriptstyle{2J \choose J+m}} f_{-J}^{(J)}
A^{J+m} =\sqrt{\scriptstyle{2J \choose J+m}} f_{J}^{(J)} (A^{-1})^{J-m}
\label{2.13}
\eeq
The starting point of the quantization is  to replace Eqs.\ref{2.8}
by their quantum counterparts, so that we now have
\begin{equation}
\Bigl [\vartheta'_1(\sigma_1),\vartheta'_1(\sigma_2) \Bigr ]=
\Bigl[\vartheta'_2(\sigma_1),\vartheta'_2(\sigma_2) \Bigr ]
=2\pi i\,  \delta'(\sigma_1-\sigma_2)
\label{2.14}
\end{equation}
It was shown in ref.\cite{GN3} that $\vartheta_1$ and $\vartheta_2$
are related by a complicated canonical transformation;
however, the relation between the zero modes is simple:
\beq
p_0^{(1)}=-p_0^{(2)} \quad ,\quad \bar p_0^{(1)}=-\bar p_0^{(2)}.
\label{zeromodes}
\eeq
Instead of $p_0^{(1)}$ (or $p_0^{(2)}$) it will be more convenient to
work with the rescaled zero mode
\beq
\varpi :=ip_0^{(1)}\sqrt{2\pi\over h}
\label{defomega}
\eeq
with $h$ defined in terms of the central charge $C$ by
\beq
h={\pi \over 12}(C-13 -\sqrt{(C-25)(C-1)})
\label{defh}
\eeq
The  parameter $h$ which is  the deformation parameter
of $sl(2)$, is also in effect the  Planck constant of the quantum Liouville
theory.

\section{The case of a single screening charge}
\subsection{The braiding of the holomorphic components}
Starting from the representation
Eq.\ref{2.13}, and following the method of ref.\cite{GS1},  we
construct the
quantum equivalents of the fields $f_m^{(J)}$. In the earlier
papers they have been noted $V_m^{(J)}$, $\Vt_m^{(J)}$, or
$U_m^{(J)}$, depending upon the normalization chosen. They are periodic
up to a multiplicative constant and thus can be considered as Bloch waves.
On the other hand, there is also a second basis of chiral operators
$\xi_M^{(J)}$,
which are by construction explicitly covariant under the quantum group
\cite{G1}\cite{CGR2}, and related to the Bloch wave fields by a linear
transformation. In the present article, we will concentrate on the
Bloch wave basis; the discussion of the $\xi_M^{(J)}$ fields for continous
spins will be carried out elsewhere.
The construction of the Bloch wave vertex operators and their exchange
algebra
 was essentially displayed
in ref.\cite{GS1}, we go through it again as a preparation for the
case of two screening charges, and to
make some points which were left out before for brevity.
As discussed above, we consider the semi-infinite families of Bloch wave
operators with $J+m$ or $J-m$ a non-negative integer.
It turns out that there exists  a consistent operator algebra where
the two types of families
do not mix\cite{GR}. Thus
  we may  concentrate on one type, say  the case
  with $J+m=0,1,\dots$.
Then  the quantum version
of $f^{(J)}_{m}$ is  most easily obtained
from the quantum versions of (the
left equality in) Eq.\ref{2.13}, and  of Eq.\ref{2.11}.
According to Eq.\ref{2.10}, this leads to quantum expressions in terms
of $\vartheta_1 $ --- note that the other case ( $J-m$ integer)  may
be obtained by the replacement
$\vartheta_1 \leftrightarrow \vartheta_2$ everywhere (cf. also section
\ref{theta1theta2inv}).

To begin with, the factor $f^{(J)}_{-J}$
is replaced by
 the normal-ordered exponential
\beq
f_{-J}^{(J)}\bigr |_{\hbox {\scriptsize qu}}
\equiv U_{-J}^{(J)}=
: e^{2J\sqrt{h/2\pi}\vartheta_1}:
\label{3.0}
\eeq
The parameter $h$ is that of Eq.\ref{defh}.
The change of the coefficient in the exponential is such that this field
has conformal weight
\begin{equation}
\Delta_J=-J -{h\over \pi}J (J +1).
\label{2.16}
\end{equation}
If  $2J$ is a positive integer, this coincides with Kac's formula,
and $U_{-J}^{(J)}$ is a
$(1, 2J+1)$ primary in the BPZ classification.
 Now let us turn to the second factor $A^{J+m}$ appearing on the
left equation of
Eq.\ref{2.13}.
As already used in \cite{GS1}, the classical
expression
Eq.\ref{2.11}  for $A$
 has a rather simple quantum generalization, which we
will denote by  $S$ to signify that it is
is a primary field of dimension zero (``screening charge''),
namely\cite{LuS}
\beq
 S(\sigma)= e^{2ih(\varpi+1)}
 \int _0 ^{\sigma} d\rho U_{1}^{(-1)}(\rho)+
 \int _\sigma  ^{2\pi } d\rho U_{1}^{(-1)}(\rho)
\label{3.1}
\eeq
Apart from an overall change of normalization ---
removal of the denominator ---  and the introduction of normal orderings,
 the only change
 consists in the replacement
$\varpi \to \varpi+1$ in the prefactor of the first integral.
 The quantum formula is such that $S$ is
periodic up to a multiplicative factor
\beq
S(\sigma+2\pi)=e^{2ih(\varpi+1)}S(\sigma).
\label{3.2}
\eeq
This is the quantum version of the left equation in Eq.\ref{2.9}.
The basic primary field of the Coulomb gas picture is now
defined as
\beq
U_{m}^{(J)}(\sigma)=  U_{-J}^{(J)}(\sigma) [S(\sigma)]^{J+m}
\label{3.3}
\end{equation}
which is the quantum version of the first equality in Eq.\ref{2.13}.
The product of operators at the same point implied in Eq.\ref{3.3}
exists for small enough $h$ (more on this  below).
Since $S$ is a screening operator, the conformal dimension of $U_m^{(J)}$
 agrees
with Eq.\ref{2.16}. Furthermore,
one easily verifies that
\begin{equation}
U_m^{(J)} \varpi = (\varpi+2m) U_m^{(J)}
\label{3.4}
\end{equation}
Here we are assuming $\varpi$ to be real, as is appropriate in
the socalled
elliptic sector of the theory (cf. section \ref{liouexp}). Also
in the rest of the paper we will concentrate on this case, if not
indicated otherwise. It is the case which appears to be directly
related to (tree level) amplitudes in $c\le 1$ string theory\cite{G5}.
The normalization of the $U_m^{(J)}$ operators is given by
\beq
<\varpi |U^{(J)}_m |\varpi+2m>=I^{(J)}_m(\varpi)
\label{3.5}
\eeq
and $I^{(J)}_m(\varpi)$ is computed in appendix
\ref{normalization integrals} to be\footnote{see ref.\cite{Fe} for a
similar calculation, applied to the degenerate case}
\[
I_m^{(J)}(\varpi )=
\left( 2 \pi \Gamma(1+{h\over\pi}) \right )^{J+m}
e^{ ih(J+m)(\varpi-J+m)}
\]
\beq
\prod_{\ell=1}^{J+m}
{\Gamma[1+(2J-\ell+1)h/\pi] \over \Gamma[1+\ell h/\pi]
 \Gamma[1-(\varpi+2m-\ell)h/\pi]
\Gamma[1+(\varpi+\ell)h/\pi] }.
\label{3.6}
\eeq
This formula illustrates  an important
point to be made about the integral representation Eq.\ref{3.3}.
For small enough $h$, the arguments of the gamma functions are
all positive,
and this corresponds to the domain where the integral
representation is
convergent. When $h$ increases, divergences
 appear. However, Eq.\ref{3.6}
continues to make sense beyond the poles by the usual
analytic continuation
of the Gamma function. As is well known\cite{GN4}, the continuation
of the ground state expectation value $I_m^{(J)}(\varpi )$ defines the
continuation of the operator $U_m^{(J)}$ itself. Thus, $U_m^{(J)}$
is related to  the normalized operator $V_m^{(J)}$ with
$\langle \varpi |V_m^{(J)}(\sigma =0) |\varpi +2m\rangle =1$
(introduced already in ref.\cite{CGR1} for half-integer positive $J$)
by
\beq
U^{(J)}_m =I^{(J)}_m(\varpi) V^{(J)}_m \quad .
\label{3.7}
\eeq
Note that $U_{-J}^{(J)}\equiv V_{-J}^{(J)}$.

We now come to
 the braiding  algebra of the fields
$U_m^{(J)}$. For half-integer positive spin --- the case
corresponding to Kac's table --- it is well known that
the braiding of the $U_m^{(J)}$ or $V_m^{(J)}$ is essentially given
by a q-$6j$-symbol, and the explicit formulae were
determined in ref.\cite{CGR1} (The general result is summarized in
appendix A).
In ref.\cite{GS1},  this result
was extended to arbitrary $J$.
We will recall some basic points of the derivation
 that will be useful later on.
The braiding relation takes the form
\beq
U_m^{(J)}(\sigma) U_m^{(J')}(\sigma')=
\sum_{m_1,m_2} R_U(J,J';\varpi)^{m_2 m_1}_{m\phantom{_2} m'\phantom{_1}}
U_{m_2}^{(J')}(\sigma') U_{m_1}^{(J)}(\sigma)
 \label{3.10}.
\eeq
We only deal with the case $\pi >\sigma '>\sigma >0$ explicitly. The other
cases are deduced from the present one in the standard way.
The sums extend over non-negative integer $J+m_1$ resp. $J'+m_2$
with the condition
\begin{equation}
m_1+m_2=m+m'=:m_{12}.
\label{3.11}
\end{equation}
Since one considers the braiding at
equal $\tau$ one can let $\tau=0$ once and for all.
As there are no  null-vector
decoupling equations for continuous $J$,
the derivation of Eq.\ref{3.10}   relies  exclusively on the free
field techniques summarized in the previous section.
The basic point of our argument
is that
 the exchange of two $U_m^{(J)}$
operators can be mapped into an equivalent problem in one-dimensional
quantum mechanics, and becomes just finite-dimensional linear algebra.
In  view of Eqs.\ref{3.1}, \ref{3.3},
 the essential observation
is  that one only needs the braiding relations of
$U_{-J}^{(J)}\equiv V_{-J}^{(J)}$ operators which are normal ordered
exponentials
(``tachyon operators''). One has
\begin{equation}
V_{-J}^{(J)}(\sigma )V_{-J'}^{(J')}(\sigma ')=
e^{-i2JJ' h \epsilon (\sigma -\sigma')}
V_{-J'}^{(J')}(\sigma ' )V_{-J}^{(J)}(\sigma )
\label{3.12}
\end{equation}
where $\epsilon (\sigma -\sigma')$ is the sign of $\sigma -\sigma'$.
This means that when commuting the tachyon operators in
$U_{m'}^{(J')}(\sigma ')$ through those of
$U_{m}^{(J)}(\sigma )$, one only encounters phase factors of the form
$e^{\pm i2 \alpha \beta h } $ resp. $e^{\pm  6i \alpha \beta h } $,
with $\alpha$ equal to $J$ or $-1$, $\beta$ equal to $J'$ or $-1$,
since  we take
$\sigma, \sigma' \in  [0,\pi]$. Hence we are led to decompose the
integrals defining the screening charges S into pieces which commute
with each other and  with $V_{-J}^{(J)}(\sigma)$, $V_{-J'}^{(J')}
(\sigma ')$ up to one of the above phase factors. We consider
explicitly only the case $0< \sigma < \sigma ' < \pi$ and write
\[
S(\sigma )\phantom{'} = S_{\sigma \sigma '} + S_\Delta,
\quad
S(\sigma ')= S_{\sigma \sigma '} + k(\varpi)S_\Delta \equiv
S_{\sigma \sigma '} + \tilde {S}_\Delta,
\]
\[
S_{\sigma \sigma '}:= k(\varpi)
\int_0^\sigma V_{1}^{(-1)}(\rho )
d\rho + \int_{\sigma '}^{2\pi}V_{1}^{(-1)}(\rho )
d\rho,
\]
\begin{equation}
S_\Delta := \int_\sigma^{\sigma '}V_{1}^{(-1)}(\rho )
d\rho,
\quad
k(\varpi):= e^{2ih(\varpi+1)}
\label{3.13}
\end{equation}
Using Eq.\ref{3.12}, we then get the following simple algebra for
$S_{\sigma \sigma '}, S_\Delta , \tilde {S}_\Delta$:
\medskip
\begin{equation}
S_{\sigma \sigma '}S_\Delta =q^{-2}S_\Delta S_{\sigma \sigma '},
\quad
S_{\sigma \sigma '}\tilde {S}_\Delta =q^2 \tilde {S}_\Delta
S_{\sigma \sigma '},
\quad
S_\Delta \tilde {S}_\Delta =q^4 \tilde {S}_\Delta S_\Delta,
\label{3.14}
\end{equation}
and their commutation properties with
$V_{-J}^{(J)}(\sigma ),V_{-J'}^{(J')}(\sigma ' )$ are given by
 \begin{equation}
\begin{array}{lll}
&V_{-J}^{(J)}(\sigma) S_{\sigma \sigma '}=q^{-2J}S_{\sigma \sigma'}
V_{-J}^{(J)}(\sigma ),
&V_{-J'}^{(J')}(\sigma')S_{\sigma \sigma '}= q^{-2J'} S_{\sigma \sigma'}
V_{-J'}^{(J')}(\sigma'),
\nonumber\\
&V_{-J}^{(J)}(\sigma)S_\Delta  = q^{-2J}S_\Delta V_{-J}^{(J)}(\sigma ),
&V_{-J}^{(J)}(\sigma) \tilde {S}_\Delta \  =q^{-6J} \tilde {S}_\Delta
V_{-J}^{(J)}(\sigma),
\nonumber \\
&V_{-J'}^{(J')}(\sigma')S_\Delta = q^{2J'} S_\Delta V_{-J'}^{(J')}
(\sigma '),
&V_{-J'}^{(J')}(\sigma')\tilde {S}_\Delta = q^{-2J'} \tilde {S}_\Delta
V_{-J'}^{(J')}(\sigma').
\end{array}
\label{3.15}
\end{equation}
Finally, all three screening pieces obviously shift the zero mode in the
same way:
\begin{equation}
 \left. \begin{array}{ccc}
S_{\sigma \sigma '}\nonumber \\ S_\Delta \nonumber \\ \tilde {S}_\Delta
\end{array} \right \} \varpi
= (\varpi +2) \left \{ \begin{array}{ccc}
S_{\sigma \sigma '}\nonumber \\ S_\Delta \nonumber \\ \tilde {S}_\Delta
\end{array} \right. .
\label{3.16}
\end{equation}
Using Eqs.\ref{3.15} we can commute $\Vt_{-J}^{(J)}(\sigma  )$ and
$\Vt_{-J'}^{(J')}(\sigma  ')$ to the left on both sides of  Eq.\ref{3.10},
so that they can be cancelled. Then we are left with
\[
(q^{-2J'}S_\Delta + q^{2J'}S_{\sigma \sigma '} )^{J+m}
(\tilde {S}_\Delta +S_{\sigma \sigma '})^{J'+m'}q^{2JJ'}=
\]
\begin{equation}
\sum_{m_1,m_2} R(J,J';\varpi + 2(J+J'))_{m_{\phantom{2}}m'}^{m_2 m_1}
(q^{2J} S_{\sigma \sigma '} + q^{6J}\tilde {S}_\Delta )^{J'+m_2}
(S_{\sigma \sigma '} + S_\Delta )^{J+m_1}
\label{3.17}
\end{equation}
It is apparent from this equation that the braiding problem of the
$U_m^{(J)}$ operators is governed by the Heisenberg-like algebra
Eq.\ref{3.14}, characteristic of one-dimensional
quantum mechanics. We will  proceed using  the following simple representation
of the algebra Eq.\ref{3.14} in terms of one-dimensional
quantum mechanics ( $y$ and $y'$ are arbitrary complex numbers):
\begin{equation}
S_{\sigma \sigma '}=y' e^{2Q}, \quad
S_\Delta =y e^{2Q-P}, \quad  \tilde {S}_\Delta =
y e^{2Q+P},
 \qquad [Q,P]=ih  .
\label{3.18}
\end{equation}
The third  relation in Eq.\ref{3.18} follows from
the second one in view of
$\tilde {S}_\Delta =k(\varpi )S_\Delta$ (cf. Eq.\ref{3.13}).
	This means we are
identifying here $P \equiv ih\varpi $ with the zero mode of the original
problem. Using $e^{2Q+cP} = e^{cP} e^{2Q} q^c$ we can commute all factors
$e^{2Q}$ to the right on both sides of Eq.\ref{3.17} and then cancel them.
This leaves us with
\[
q^{2JJ'}
\prod_{s=1}^{J+m}(y'q^{2J'} +yq^{-(\varpi -2J+2s-1)})
\prod_{t=1}^{J'+m'} (y'+yq^{\varpi -2J'+2m+2t-1})=
\]
\begin{equation}
\sum_{m_1} R_U(J,J';\varpi)_{m_{\phantom {2}} m'}^{m_2 m_1}
\prod_{t=1}^{J'+m_2}(y'q^{2J} +yq^{\varpi +4J-2J'+2t-1})
\prod_{s=1}^{J+m_1} (y'+yq^{-(\varpi -2J+2m_2 +2s -1)})
\label{3.19}
\end{equation}
where we have shifted back $\varpi +2(J+J') \rightarrow \varpi$
compared to Eq.\ref{3.17}.
Since the overall scaling $ y\rightarrow \lambda y,
y' \rightarrow \lambda y'$ only gives back Eq.\ref{3.11}, we can set $y'=1$.

The solution of these equations, which was
 derived in ref.\cite{GS1}, will be
cast under the convenient form
\beq
R_U(J,J',\varpi)_{m_{\phantom{2}} m'}^{m_2 m_1}=
e^{-i\pi (\Delta_c+\Delta_b
-\Delta_e-\Delta_f)}
{\kappa_{ab}^e \kappa_{de}^c\over \kappa_{db}^f \kappa_{af}^c}
\left\{ ^{a}_{d}\,^{b}_{c}
\right. \left |^{e}_{f}\right\}
\label{3.20}
\eeq
where $\left\{ ^{a}_{d}\,^{b}_{c}
\right. \left |^{e}_{f}\right\}$ is the q-$6j$-symbol generalized
to continous spins, with arguments
$$
a=J,\quad b=x+m+m',\quad c=x\equiv (\varpi-\varpi_0)/2
$$
\beq
d=J',\quad e=x+m_2,\quad f=x+m,
\label{3.21}
\eeq
and the coefficients $\kappa_{J_1 J_2}^{J_{12}}$ are given by
\footnote{We follow the prescription of refs.\cite{G5},\cite{GR} for
the definition of the square roots (cf. also below Eq.\ref{defg}).
In particular, the square root in Eq.\ref{3.9} is to be understood
as the product resp. quotient of the square roots of the individual
factors consisting of a single q-number.}
$$
\kappa_{J_1 J_2}^{J_{12}}=
\left ({ he^{-i(h+\pi)}
\over 2\pi \Gamma(1+h/\pi) \sin h}\right )^{J_1+J_2-J_{12}}
e^{ih (J_1+J_2-J_{12})( J_1-J_2-J_{12})} \times
$$
\beq
\prod_{k=1}^{J_1+J_2-J_{12}}
\sqrt{ \lfloor 1+2J_1-k\rfloor \over
\lfloor k \rfloor \, \lfloor 1+2J_2-k \rfloor\,
\lfloor -(1+2J_{12}+k) \rfloor }.
\label{3.9}
\eeq
Recall that we let $\lfloor x \rfloor =\sin(hx)/\sin h$ in general.
The  last equation  makes sense
for arbitrary $J_1,J_2, J_{12}$ such that $J_1+J_2-J_{12}$ is a
non-negative integer.
The r.h.s.of Eq.\ref{3.20}
 may be expressed in terms of q-hypergeometric functions
by the formula
\[
{\kappa_{ab}^e \kappa_{de}^c\over \kappa_{db}^f \kappa_{af}^c}
\left\{ ^{a}_{d}\,^{b}_{c}
\right. \left |^{e}_{f}\right\}
=
q^{2dn-2an_2}\lfloor 2e+1\rfloor
{\lfloor \phi \rfloor_{n_1}
 \lfloor 1-\epsilon-n_1\rfloor_{n_1} \over
\lfloor n_1 \rfloor \! !
\lfloor \beta-\epsilon-n'-n \rfloor_{n+n'+1}
 }\times
\]
\beq
\times\lfloor \rho\rfloor_{n'}
\lfloor 1-\beta\rfloor_{n-n_1}
\>_4 F_3\left (^{\alpha,}_{\epsilon,} \,\,
^{\beta ,}_{\phi ,} \,\, ^{-n_1 ,}_
{\rho } \,\, ^{-n'} \ ;q,1 \right )
\label{3.22}
\eeq
with\hfill
$$
\alpha=-a-c+f,\quad \beta=-c-d+e,
$$
$$
n_1=a+b-e, \quad n_2=d+e-c,\quad  n=f+a-c \quad n'=b+d-f;
$$
\beq
\epsilon=-(a+b+c+d+1), \quad
\phi=1+n-n_1, \quad
\rho=e+f-a-d+1.
\label{3.23}
\eeq
We have  defined, as in the previous work along the same line,
\[
_4 F_3 \left (^{a,}_{e,} \, ^{b,}_{f,} \,
^{c,}_{g} \, ^{d} ;q,\rho \right )=
\sum_{n=0}^\infty {\lfloor a\rfloor_n \lfloor b\rfloor_n
\lfloor c\rfloor_n \lfloor d\rfloor_n \over \lfloor e\rfloor_n
\lfloor f\rfloor_n \lfloor g\rfloor_n \lfloor n\rfloor !} \ \rho^n ,
\]
\beq
\lfloor a\rfloor_n := \lfloor a\rfloor \lfloor a+1\rfloor \cdots
\lfloor a+n-1\rfloor ,
\quad \lfloor a\rfloor_0 :=1.
\label{3.25}
\eeq
 In  Eq.\ref{3.22}, the prefactor involves
 products of the type just recalled with
indices
\beq
n_1=J+m_1, \quad
n_2=J+m_2, \quad
n=J+m, \quad
n'=J+m'.
\label{3.24}
\eeq
Since  they are equal to the screening numbers,
they  are positive integers. Thus  Eq.\ref{3.22}
makes sense for arbitrary spins provided the screening numbers are
integers, and is the appropriate generalization.

The method used to derive Eq.\ref{3.20} was to transform\footnote{This
transformation formula is derived in ref.\cite{GR}.} the
hypergeometric function into another one such that
the desired relations
Eq.\ref{3.19}
follow from the orthogonality relation of the associated Askey-Wilson
polynomials. In this connection, let us  note that a
simple reshuffling of the parameters of the latter form
 allows to verify that
the usual orthogonality relations of the 6-j symbols extend
to our case. One has,  in general,
\beq
\sum_{J_{23}}
\left\{ ^{J_1 }_{J_3 }
\> ^{J_2 } _{ J_{123}}
\right.
\left |^{ J_{12}}
_{J_{23}}\right\}
\left\{ ^{J_1 }_{J_3 }
\> ^{J_2 } _{ J_{123}}
\right.
\left |^{ K_{12}}
_{J_{23}}\right\} =\delta_{J_{12}-K_{12}}
\label{orth}
\eeq
where the $J$'s are arbitrary except for the constraint that
  the screening numbers
$$
n_1=J_1+J_2-J_{12}, \quad n_2=J_3+J_{12}-J_{123}, \quad
n=J_1+J_{23}-J_{123},\quad n'=J_2+J_3-J_{23},
$$
\beq
\tilde n_1=J_1+J_2-K_{12}, \quad \tilde n_2=J_3+K_{12}-J_{123},
\label{3.26}
\eeq
are positive integers. These  conditions fix the range of summation over
$J_{23}$.

The basis $U_m^{(J)}$, apart from its manageability, has another practical
virtue: Its braiding (and fusion) properties  are given in a form
which involves no square roots, but
only (q-deformed) rational functions, and no phase
ambiguities can arise. On the other hand,
from the quantum group point of view
it is more natural to consider a basis where
the braiding (and fusion) is given exclusively
in terms of the $6j$-symbol (the latter does however
involve square roots).
For this purpose, the authors of ref.\cite{CGR1} introduced the fields
$\Vt_m^{(J)}$, which are defined by
\beq
\Vt_m^{(J)}=g_{J,x+m}^{x} V_m^{(J)}.
\label{3.27}
\eeq
As before we let  $x=(\varpi-\varpi_0)/2$. The coupling constants
$g$ are defined by
$$
g_{J, \, x +m}^{x}=
\left ( {h\over\pi}\right)
^ {J+m}
\prod_{k=1}^{J+m}
\sqrt{
F[1+(2J-k+1)h/\pi]}
\sqrt{F[(\varpi+2m-k)h/\pi]} \times
$$
\beq
\prod_{k=1}^{J+m} \sqrt{ F[-(\varpi+k)h/\pi]} \left  /
\sqrt{F[1+kh/\pi]}\right.
\label{defg}
\eeq
where, as usual,  $F(z):=\Gamma(z)/\Gamma(1-z)$. The treatment of the
square roots requires some care. We follow the prescription
of ref.\cite{G5}
also used in ref.\cite{GR}. Eq.\ref{defg} immediately extends to
the case of non integer $J$, as $J+m$ remains a positive integer.
We then have ($m_{12}:=m_1+m_2=m+m'$)
\[
\Vt_m^{(J)}(\sigma)\Vt_{m'}^{(J')}(\sigma')=
e^{-i\pi(\Delta_{x} +\Delta_{x+m_{12}}
-\Delta_{x+m} -\Delta_{x+m_2})}\times
\]
\beq
\sum_{m_1, m_2}
\left\{ ^{J\quad }_{J'\quad }
\> ^{x +m_{12}}
_{ x\phantom{+m_{12}}}
\right.
\left |^{\quad  x+m_2}
_{\quad  x+m}\right\}
\Vt_{m_2}^{(J')}(\sigma')\Vt_{m_1}^{(J)}(\sigma)
\label{3.29}
\eeq
The relation with $U^{(J)}_m$ is given by
\beq
U^{(J)}_m ={I^{(J)}_m(\varpi)\over
g_{J, \, x +m}^{x}} \Vt^{(J)}_m
\equiv {1\over \kappa_{J, \, x +m}^{x}}
\Vt^{(J)}_m.
\label{3.8}
\eeq
The $\kappa$ coefficient should , of course, be given by Eq.\ref{3.9}.
This is checked in appendix \ref{normalization integrals}.
Concerning the right-moving
modes, the braiding algebra is given by
\[
\Vtb_m^{(J)}(\sigma)\Vtb_{m'}^{(J')}(\sigma')=
e^{i\pi(\Delta_{\xb} +\Delta_{\xb+m_{12}}
-\Delta_{ \xb+m} -\Delta_{\xb+m_2})}\times
\]
\beq
\sum_{m_1, m_2}
\left\{ ^{J_1\quad }_{J_2\quad }
\> ^{\xb +m_{12}}
_{ \xb\phantom{+m_{12}}}
\right.
\left |^{\quad  \xb+m_2}
_{\quad  \xb+m}\right\}
\Vtb_{m_2}^{(J')}(\sigma')\Vtb_{m_1}^{(J)}(\sigma)
\label{Vtbar}
\eeq
where we let $\xb=(\varpib-\varpi_0)/2$.
The only difference with Eq.\ref{3.29} is the change of sign of the
phase factor. This may be verified  by redoing the whole derivation.
In ref\cite{G5}, it was remarked that the right-mover braiding matrix is
deduced from the left-mover one by changing $i=\sqrt{-1}$ into $-i$, since
this correctly changes the orientation of the complex plane.
This complex conjugation is most easily performed using the $U$ fields,
since the braiding matrix Eq.\ref{3.20} is real apart from the
first phase factor. For the $\Vt$ fields,
there is a
slight subtlety
 related again
to the appearance of the redundant
square roots in Eq.\ref{3.9}. The  correct rule is to
take the same definition for the square roots for left and right movers.
Thus  the right-moving
coupling constant
$\bar g^{\xb}_
{J,\xb+m}$
is  given
by the {\it same} expression Eq.\ref{defg}, not its complex conjugate.
The same
prescription should be followed for the roots appearing in
${\overline \kappa}^{\xb}_{J,\xb+m}$,
while taking the usual complex conjugate for the phase factor
appearing in front
of the product in Eq.\ref{3.9}. Note that $\varpi$ is always to be
	treated as
real formally in this context, even in the hyperbolic sector where it is
actually purely imaginary (cf. below).

\section{Solving of the Liouville quantum dynamics}
\subsection{The Liouville exponential}
\label{liouexp}
First, let us note   that, as  $h$ is real
 in the weak-coupling regime,
 the hermiticity of energy-momentum
allows for $\varpi,\varpib$ real or purely imaginary, corresponding to
the elliptic resp. hyperbolic sector of the theory\cite{JKM}
(see also \cite{GN2} for the case of open boundary conditions).
In the former case, which we consider in this paper, we will see that
the locality conditions are fulfilled if
\begin{equation}
\varpi-\varpib=k\pi/h,\qquad\qquad k\in {\bf Z}
\label{condx}
\end{equation}
Eq.\ref{condx} has an immediate interpretation as the natural generalization
of the classical boundary conditions\cite{LuS}.
Moreover we will show that the appropriate definition of the
Liouville exponential  for arbitrary $J$ is
\beq
e^{\textstyle -J\alpha_-\Phi(\sigma, \tau )}=
\sum _{m=-J}^{\infty} \mu_0^{J+m}
\Vt_m^{(J)}(u)\,
{\overline \Vt_{m}^{(J)}}(v) \>
\label{3.31}
\eeq
where $\alpha_- =2\sqrt{h/2\pi}$ is the screening charge.
The constant $\mu_0^{J+m}$ will not be fixed by braiding or fusion.
It will be determined below when we derive  the field equations.
\subsubsection{Locality}
Let us now check locality.
In the approach of refs.\cite{LuS}\cite{G5}, one takes
the zero modes of the left-moving and right-moving Liouville modes
to commute, so that $U$ and ${\overline U}$ commute. However, operators
involving both chiralities should be applied only to states
fulfilling Eq.\ref{condx}, and conserve this condition. This is why we must
have
$\bar m=m$ in Eq.\ref{3.31}.\footnote{The situation is quite different
in the strong coupling theory, see e.g. ref.\cite{GR}.}
Next we observe that if Eq.\ref{condx}
is valid,
 \beq
{\overline R}_{\Vtb}(J,J';{\overline\varpi})
^{\bar m_2 \bar m_1}_{\bar m_{\phantom{2}} \bar m'}=
{\overline R}_{\Vtb}(J,J';\varpi)^{\bar m_2 \bar m_1}_{\bar m_{\phantom{2}}
\bar m'}
\label{omegabar}
\eeq
as can be verified easily. The same is  true for ${\overline R}_{\overline U}$.
 Thus we have
\[
\Vtb_m^{(J)}(\sigma)\Vtb_{m'}^{(J')}(\sigma')=
e^{+i\pi(\Delta_{x} +\Delta_{x+m_{12}}
-\Delta_{x+m} -\Delta_{x+m_2})}\times
\]
\beq
\sum_{m_1, m_2}
\left\{ ^{J_1\quad }_{J_2\quad }
\> ^{x+m_{12}}
_{ \quad \quad x}
\right.
\left |^{\quad  x +m_2}
_{\quad  x+m}\right\}
\Vtb_{m_2}^{(J')}(\sigma')\Vtb_{m_1}^{(J)}(\sigma)
\label{Vtbar2}
\eeq
Then,
according to Eqs.\ref{3.29}, \ref{Vtbar2} we get
$$
e^{\textstyle -J\alpha_-\Phi(\sigma, \tau )}
e^{\textstyle -J'\alpha_-\Phi(\sigma', \tau )}=
$$
$$
\sum_{m, m'}
\sum_{m_1, m_2; \mb_1, \mb_2}e^{ih(\mb_2-m_2)(\mb_2-m_2 +\varpi)}
\left\{ ^{J\quad }_{J'\quad }
\> ^{x +m_1+m_2}
_{x\phantom{+m_1+m_2}}
\right.
\left |^{ x+m_2}
_{ x+m}\right\}
\left\{ ^{J\quad }_{J'\quad }
\> ^{x+m_1+m_2}
_{x\phantom{+m_1+m_2}}
\right.
\left |^{x +\mb_2}
_{x+m}\right\}
$$
\beq
\times  \mu_0^{J+m+J'+m'}
\left. \Vt_{m_2}^{(J')}(u') \Vt_{m_1}^{(J)}(u)\,
{\overline \Vt_{m_2}^{(J')}}(v') {\overline \Vt_{m_1}^{(J)}}(v)
 \right |_{m_1+m_2=m+m' \atop
{\overline m}_1+{\overline m}_2=m+m'},
\label{braidexp}
\eeq
  One first sums  over $m$, with fixed
$m_{12}=m+m'$. This precisely corresponds to the
 summation over $J_{23}$ in Eq.\ref{3.26}.
Thus only $m_2=\mb_2$ contributes.
This gives immediately
\beq
e^{\textstyle -J_1\alpha_-\Phi(\sigma_1, \tau )}
e^{\textstyle -J_2\alpha_-\Phi(\sigma_2, \tau )}=
e^{\textstyle -J_2\alpha_-\Phi(\sigma_2, \tau )}
e^{\textstyle -J_1\alpha_-\Phi(\sigma_1, \tau )},
\label{loc}
\eeq
and the Liouville exponential is local for arbitrary $J$.
We remark that Eq.\ref{condx} is not only sufficient, but also
necessary for locality, as was observed in ref.\cite{LuS} for the
special case $J=1/2$.
\subsubsection{Closure by fusion}
In the preceding analysis, we have discussed only the braiding properties
of the chiral fields resp. the Liouville exponentials. However, according
to the general Moore-Seiberg formalism\cite{MS}, fusion (in the sense of the
full operator product) and braiding are not independent. Assuming
the validity of the Moore-Seiberg relation between fusion and braiding matrix,
we then obtain immediately
that the fusion of the
 $\Vt$ fields should be given by (cf. also ref.\cite{GR})
$$
\Vt^{(J_1)}_{m_1}(z_1) \Vt^{(J_2)}_{m_2}(z_2) =
\sum _{J_{12}= -m_1-m_2} ^{J_1+J_2}
\left\{ ^{ J_1}_{x +m_1+m_2}\,
\> ^{ J_2}_{x}
\right.
\left |^{ J_{12}}_{x+m_1}\right\}
\times
$$
\beq
\sum _{\{\nu_{12}\}}
\Vt ^{(J_{12},\{\nu_{12}\})}_{m_1+m_2}(z_2)
<\!\varpi_{J_{12}},{\{\nu_{12}\}} \vert
\Vt ^{(J_1)}_{J_2-J_{12}}(z_1-z_2) \vert \varpi_{J_2} \! >.
\label{fus1}
\eeq
In Eq.\ref{fus1} we have changed variables by letting
$z=e^{i(\tau+\sigma)}$, $\zb=e^{i(\tau-\sigma)}$ (recall that we are using
Minkowski world-sheet variables).\footnote{with Euclidean variables, this would
mean that we change to the sphere.} The only difference to the positive
half-integer spin case, which was completely analyzed in ref.\cite{CGR1},
is that
the $J_{12}$ -sum now extends to $-m_1-m_2$ instead of $|J_1-J_2|$.
Indeed, the positivity of the screening numbers appearing in the braiding
matrix leads via the Moore-Seiberg relation to the positivity of the
screening numbers $n_1=J_1+m_1,\  n_2=J_2+m_2,\ p_{1,2}=J_1+J_2-J_{12},
 \ n=J_{12}+m_1+m_2$ of the fusion matrix. In
ref.\cite{GR}, it has been shown that the generalized $6j$-symbol
of Eq.\ref{3.20}, together with the positivity condition for the
screening charges, fulfills all the necessary identities for the
polynomial equations to be valid with continous spins. This provides a strong
argument that the fusion matrix of Eq.\ref{fus1} is indeed the correct one,
even though we have not attempted to derive it directly as we did for the
braiding.
Making use of the analogous equation for the bar components, one sees that the
operator-product expansion of Liouville exponentials may be written as
$$
e^{\textstyle -J_1\alpha_-\Phi(z_1, \zb_1 )}
e^{\textstyle -J_2\alpha_-\Phi(z_2, \zb_2 )}=
$$
$$
\sum_{m_1,m_2} \sum _{J_{12}, \Jb_{12}}
\left\{ ^{ J_1}_{x +m_1+m_2}\,
\> ^{ J_2}_{x}
\right.
\left |^{ J_{12}}_{x+m_1}\right\}
\left\{ ^{ J_1}_{\xb +m_1+m_2}\,
\> ^{ J_2}_{\xb}
\right.
\left |^{ \Jb_{12}}_{\xb+m_1}\right\}\times
$$
$$
\mu_0^{J_1+m_1+J_2+m_2}
\sum _{\{\nu_{12}\}, \{\nub_{12}\} }
\Vt ^{(J_{12},\{\nu_{12}\})}_{m_1+m_2}(z_2)
\Vtb ^{(\Jb_{12},\{\nub_{12}\})}_{m_1+m_2}(\zb_2)\times
$$
\beq
<\!\varpi_{J_{12}},{\{\nu_{12}\}} \vert
\Vt ^{(J_1)}_{J_2-J_{12}}(z_1-z_2) \vert \varpi_{J_2} \! >
<\!\varpib_{\Jb_{12}},{\{\nub_{12}\}} \vert
\Vtb ^{(J_1)}_{J_2-\Jb_{12}}(\zb_1-\zb_2) \vert \varpib_{J_2} \! >
\label{fus2}
\eeq
It follows from condition Eq.\ref{condx}  that
$\left\{ ^{ J_1}_{\xb +m_1+m_2}\,
\> ^{ J_2}_{\xb}
\right.
\left |^{ \Jb_{12}}_{\xb+m_1}\right\}=
\left\{ ^{ J_1}_{x +m_1+m_2}\,
\> ^{ J_2}_{x}
\right.
\left |^{ \Jb_{12}}_{x+m_1}\right\}$. In the same way as for locality,
the summation over $m_1$ with fixed $m_1+m_2$ then
reduces to the orthogonality relation for
$6j$-symbols so that only $J_{12}=\Jb_{12}$ contributes, and one gets
$$
e^{\textstyle -J_1\alpha_-\Phi(z_1, \zb_1 )}
e^{\textstyle -J_2\alpha_-\Phi(z_2, \zb_2 )}=
$$
$$
\sum_{m, J_{12}} \sum _{\{\nu_{12}\}, \{\nub_{12}\} }
\mu_0^{J_{12}+m}
\Vt ^{(J_{12},\{\nu_{12}\})}_{m}(z_2)
\Vtb ^{(J_{12},\{\nub_{12}\})}_{m}(\zb_2) \mu_0^{J_1+J_2-J_{12}}\times
$$
\beq
<\!\varpi_{J_{12}},{\{\nu_{12}\}} \vert
\Vt ^{(J_1)}_{J_2-J_{12}}(z_1-z_2) \vert \varpi_{J_2} \! >
<\!\varpib_{J_{12}},{\{\nub_{12}\}} \vert
\Vtb ^{(J_1)}_{J_2-J_{12}}(\zb_1-\zb_2) \vert \varpib_{J_2} \! >.
\label{fus3}
\eeq
The second line clearly involves the descendants of the
Liouville exponentials
which we denote by
\beq
e^{\textstyle -J\alpha_-\Phi^{\{\nu\}, \{\nub\}}(z, \zb )}
\equiv
\sum_{m} \mu_0^{J+m}
\Vt ^{(J_{12},\{\nu\})}_{m}(z)
\Vtb ^{(J_{12},\{\nub\})}_{m}(\zb)
\label{fus4}
\eeq
As regards the last line, it is simply the corresponding matrix element of the
Liouville exponential. One finally gets
$$
e^{\textstyle -J_1\alpha_-\Phi(z_1, \zb_1 )}
e^{\textstyle -J_2\alpha_-\Phi(z_2, \zb_2 )}=
\sum_{ J_{12}=-m_1-m_2}^{J_1+J_2}  \sum _{\{\nu\}, \{\nub\} }
e^{\textstyle -J_{12} \alpha_-\Phi^{\{\nu\}, \{\nub\}}(z_2, \zb_2 )}\times
$$
\beq
< \!\varpi_{J_{12}}, \varpib_{J_{12}}; \{\nu\}, \{\nub\} |
e^{\textstyle -J_1\alpha_-\Phi(z_1-z_2, \zb_1-\zb_2 )} |
 \varpi_{J_2}, \varpib_{J_2} \! >.
\label{fus5}
\eeq
The notation for the matrix element should be self-explanatory
\footnote{It is implied here that charge conservation should be used for
the evaluation of the matrix element, such that only the term appearing in
Eq.\ref{fus3} survives. According to ref.\cite{Cargese}, charge conservation
actually does not hold for the 3-point functions
$\langle  \varpi'|e^{-J\alpha\Phi}(z)
|\varpi\rangle$ with continous $J$. From this point of view, the notation of
Eq.\ref{fus5}
is of course not rigorously appropriate.}.
One sees that the Liouville exponential is closed by fusion for arbitary
$J$ to all orders in the descendants.
\subsubsection{The cosmological constant revisited.}
The braiding relation is invariant under the transformation
\beq
\exp(-J\alpha_-\Phi(z, \zb )) \to T \exp(-J\alpha_-\Phi(z, \zb )) T^{-1},
\label{simtrafo}
\eeq
where $T$ is an arbitrary function of the zero modes $\varpi$ and $\varpib$.
This is
why locality does not completely determine the Liouville exponentials.
We have discussed this point in detail in ref.\cite{GS2}.
Concerning  the fusion equation, the transformation just considered
does not act on the last term on the right-hand side which  is a c-number.
The definition Eq.\ref{3.31} we have chosen is such that this term
 --- a  compact book-keeping device to handle all the descendants ---
is precisely given by the matrix elements of the Liouville exponential
itself, without any additional normalization factor. It is thus
quite natural. Note however that this choice of normalization differs from
the previous one\cite{G5}; this point is discussed
in appendix \ref{connection} and chapters \ref{theta1theta2inv},
\ref{hermiticity} below.
The only remaining ambiguity  is the arbitrariness in $\mu_0$.
Changing this parameter is  tantamount
to changing  the cosmological constant following ref.\cite{G5}. Indeed,
the fusing and
braiding relations of the $\Vt$ fields are invariant if we  make the
change $\Vt_m^{(J)}\to \mu_c^{(J+m)/2} \Vt_m^{(J)}$.
Any such change is generated by a combination of a field
redefinition ($\alpha_-\Phi\to\alpha_-\Phi-\ln \mu_c$) and a similarity
transformation of the form Eq.\ref{simtrafo}.
Thus the most general field satisfying Eqs.\ref{loc}
and \ref{fus5} is given by
\beq
e^{\textstyle -J\alpha_-\Phi(\sigma, \tau )}_{\mu_c}
=\sum _{m=-J}^{\infty}
(\mu_0 \mu_c)^{J+m}\Vt_m^{(J)}(u)\,
{\overline \Vt_{m}^{(J)}}(v)=\mu_c^{J} \mu_c^{-\varpi/2}
e^{\textstyle -J\alpha_-\Phi(\sigma, \tau )}\mu_c^{\varpi/2}
\label{muc}
\eeq
We will determine $\mu_0$ below so that it corresponds  to a
cosmological constant equal to one.
\subsubsection{Expression in terms of Coulomb-gas fields}

According to
Eqs.\ref{3.3} and \ref{3.8}, Eq.\ref{3.31} may be rewritten as
$$
e^{\textstyle -J\alpha_-\Phi(\sigma, \tau )}=
\sum _{m=-J}^{\infty}
\mu_0^{J+m} \kappa_{J, \, x +m}^{x}
\kappab _{J, \, \xb +m}^{\xb}
U^{(J)}_{m}(u) {\overline U}^{(J)}_{m}(v)
$$
It is easy to see using condition Eq.\ref{condx} that
\beq
\kappab_{J, \, \xb +m}^{\xb}= \kappab_{J, \, x +m}^{x}
\label{kappashift}
\eeq
Thus  the square roots combine pairwise
and we are left with a  rational expression.
$$
e^{\textstyle -J\alpha_-\Phi(\sigma, \tau )}=
\sum _{m=-J}^{\infty} \tilde \mu_0^{J+m} (-1)^{J+m}
\prod_{k=1}^{J+m}
{ \lfloor 1+2J-k\rfloor \over
\lfloor k \rfloor \, \lfloor \varpi+2m-k \rfloor\,
\lfloor \varpi +k \rfloor } \times
$$
\beq
V^{(J)}_{-J}(u) {\overline V}^{(J)}_{-J}(v) S^{J+m} \Sb^{J+m},
\label{Cbgas}
\eeq
where we have let
\beq
\tilde \mu_0= \mu_0 \left ({   h
\over 2\pi \Gamma(1+h/\pi) \sin h} \right )^2
\label{mu0tilde}
\eeq
Using this Coulomb-gas  expression,  together
with the mentioned orthogonality relations for Askey-Wilson polynomials,
it is then possible to  directly verify  the locality of the Liouville
exponential, without encountering any square root ambiguity.
Note that Eq.\ref{Cbgas} depends only on $\varpi$, not on $\varpib$.
On the other hand, the analysis of ref.\cite{LuS} for $J=1/2$ in the
elliptic sector, when translated to the Coulomb gas basis, gives
coefficients with an explicit dependence on $k$ of Eq.\ref{condx}.
Nevertheless, the two forms are equivalent, as they must, by means of a
basis transformation Eq.\ref{simtrafo}, hence indistinguishable from
the point of view of locality.
\subsubsection{ $\vartheta_1 \leftrightarrow \vartheta_2$
invariance}
\label{theta1theta2inv}
In section \ref{self-contained} we noted the existence of a symmetry
of the theory under the exchange of the two free fields $\vartheta_1$
under $\vartheta_2$, the residual symmetry remaining after fixing
the $SL_2({\bf C})$ invariance. On the other hand, on the quantum level
the expressions we have derived in the present paper for the Liouville
field and its exponentials are not evidently symmetric under
this exchange.
However, we must remember here that the requirement of locality really
fixed these operators only up to a similarity transformation Eq.\ref{simtrafo}
(the particular form Eq.\ref{3.31} resp. Eq.\ref{Cbgas} was only distinguished
by its simplicity and its natural behaviour under fusion). Thus
a priori we can expect $\vartheta_1\leftrightarrow\vartheta_2$ invariance
only to be valid up to a similarity transformation. As a matter of fact,
we will show  (for $J$ half-integer positive) that there exists
a transformation $T(\varpi,\varpib)$ such that
\beq
T(\varpi,\varpib) e^{-J\alpha_-\Phi}_{(1)} T^{-1}(\varpi,\varpib)=
T(-\varpi,-\varpib) e^{-J\alpha_-\Phi}_{(2)} T^{-1}(-\varpi,-\varpib).
\label{sl2inv}
\eeq
where the index (1) resp. (2) indicates the
 use of the $\vartheta_1$ resp. $\vartheta_2$ representation.
This  shows in addition that
 it is possible to choose
 particular representatives in the equivalence class of  fields defined
by Eq.\ref{simtrafo}
which are manifestly $\vartheta_1\leftrightarrow\vartheta_2$ symmetric.
To prove this
 we first observe that $\vartheta_1\leftrightarrow\vartheta_2$
takes $\varpi$ into $-\varpi$ (cf. Eq.\ref{zeromodes}), whereas  the
normalized operators $V_m^{(J)}$ behave as
\beq
V_m^{(J)} \quad \rightarrow \quad V_{-m}^{(J)}
\label{Uexchange}
\eeq
(similarly for the right-movers). The latter follows
by comparison of the conformal weights and zero mode shifts of
the  $V_m^{(J)}$ operators built from $\vartheta_1$ resp. $\vartheta_2$,
as these two properties define normalized primary fields uniquely.
For positive half-integer $J$, the summation range
in Eq.\ref{Cbgas} is $m=-J,\dots J$, hence symmetric under $m \to -m$,
and the exchange $\vartheta_1 \leftrightarrow \vartheta_2$ essentially
amounts only to a reorganization of terms. Then after commuting the
$T$ operators to the left or to the right on both sides, Eq.\ref{sl2inv}
can be solved straightforwardly. A particular solution is
\beq
T(\varpi,\varpib)=\sqrt{
{\Gamma(1-\varpi h/\pi) \Gamma(1-\varpib h/\pi)\sqrt{\lfloor \varpi\rfloor
\lfloor \varpib\rfloor} \over
\Gamma(1+\varpi)\Gamma(1+\varpib)}}\mu_0^{(\varpi+\varpib)/4}
\label{T}
\eeq
The last factor means effectively\footnote{Actually the last factor in
Eq.\ref{T} removes only $\mu_0^m$ in Eq.\ref{muc}, but the remaining
normalization constant $\mu_0^J$ plays no role here. It will become important,
 however, when we consider the equations of motion.} that we should put
 $\mu_0=1$ in Eq.\ref{3.31}
and Eq.\ref{Cbgas} (cf Eq.\ref{muc}), and so we will take $\mu_0=1$ in the
following.
For $\varpi=\varpib$, Eq.\ref{T} reduces to the transformation written in
appendix \ref{connection} to establish the connection between Eq.\ref{Cbgas}
or \ref{3.31} and the exponentials of ref.\cite{G5}.
Thus the latter
are also invariant  under the $\vartheta_1\leftrightarrow\vartheta_2$
symmetry, and the same is true for the exponential
of ref.\cite{LuS} which was constructed for arbitrary $\varpi,\varpib$.
The solution Eq.\ref{T} is unique up to the replacement $T(\varpi,\varpib)
\to T(\varpi,\varpib)T_1 (\varpi,\varpib)$, with
\beq
{ T_1(\varpi,\varpib)T_1(-\varpi-2m,-\varpib -2m)\over
T_1(\varpi+2m,\varpib+2m) T_1(-\varpi,-\varpib)}=1
\label{Tfreedom}
\eeq
Unfortunately the case of continous $J$, Eq.\ref{sl2inv} is not so
easy to analyze, as the family of operators $V_m^{(J)}$ with $J+m=0,1,2,\dots$
is no longer
invariant under the replacement $m \to -m$, and Eq.\ref{sl2inv} becomes
highly nontrivial. We leave this problem for a future publication and will
restrict also in the next subsection to the case of positive
half-integer $J$.
\subsubsection{Hermiticity}
\label{hermiticity}
Another property of the Liouville exponentials that has not yet been
discussed is hermiticity. As was worked out by Gervais and Neveu a long
time ago\cite{GN4}, the free fields possess the following behaviour
under hermitian conjugation (for brevity of notation we write only the
left-movers explicitly):
\[
\vartheta_1^{{\dag}} =\vartheta_1 \quad , \quad \vartheta_2^{\dag} =\vartheta_2
\qquad (\varpi=-\varpi^\ast)
\]
\beq
\vartheta_1^{\dag} =\vartheta_2 \quad , \quad \vartheta_2^{\dag} =\vartheta_1
\qquad (\varpi=\varpi^\ast)
\label{thetaherm}
\eeq
The first case corresponds to the hyperbolic sector of the theory,
the second to the elliptic sector which we consider here.
Consequently, one has for the vertex operators
resp. screening charges:
\[
{V_{-J}^{(J)}}^{\dag} =V_{-J}^{(J)} \ , \ {V_J^{(J)}}^{\dag} =V_J^{(J)} \ ,
\ S_{(i)}^{\dag} =S_{(i)} \quad (\varpi=-\varpi^\ast)
\]
\beq
{V_{-J}^{(J)}}^{\dag} =V_{J}^{(J)} \ , \ {V_J^{(J)}}^{\dag} =V_{-J}^{(J)} \ ,
\ S_{(i)}^{\dag} =S_{(i)} \quad (\varpi=\varpi^\ast)
\label{Uherm}
\eeq
where $S_{(i)}$ denotes the screening charge constructed from $\vartheta_{i}$.
It is then immediate to show that in the elliptic sector,
\beq
(e^{-J\alpha_-\Phi }_{(1)})^{\dag}=e^{-J\alpha_-\Phi }_{(2)} \qquad
(\varpi=\varpi^\ast)
\label{expherm2}
\eeq
Our exponentials can formally be interpreted also in the hyperbolic sector,
and fulfill there
\beq
(e^{-J\alpha_-\Phi }_{(i)})^{\dag}=e^{-J\alpha_-\Phi }_{(i)} \qquad (\varpi=
-\varpi^\ast)
\label{expherm1}
\eeq
However, their locality properties are not entirely obvious in this sector-cf.
below. Returning to the elliptic case, we note that
Eq.\ref{sl2inv} and Eq.\ref{expherm2} imply
\[
(Te_{(1)}^{-J\alpha_-\Phi}T^{-1})^{\dag} =
C (Te_{(1)}^{-J\alpha_-\Phi}T^{-1}) C^{-1} \ ,
\]
with\hfill
\beq
C(\varpi,\varpib)=T^{-1{\dag}}(\varpi,\varpib)T^{-1}(-\varpi,-\varpib)
\label{expherm3}
\eeq
Thus, hermiticity is realized only up to a similarity transformation.
In fact, in the elliptic sector there exists no similarity transformation
 $T$ at all such that
$C$ becomes trivial, even if Eq.\ref{sl2inv} is not imposed. This fact
was first observed\footnote{More precisely, it was pointed out
in the second of refs.\cite{LuS}
 that $e^{-\alpha_-\Phi/2}$ can be chosen hermitian resp. antihermitian
in certain regions of $\varpi,\varpib$ space, but $e^{-\alpha_-\Phi/2}$
cannot be consistently restricted to these regions.}
in ref.\cite{LuS} and later rediscovered in \cite{BP}.
 Nevertheless, the weaker
hermiticity property Eq.\ref{expherm3} serves almost the same purpose as
``true'' hermiticity as far as correlators of the Liouville exponentials
are concerned, as the similarity transformation $C$ cancels out up to
the contributions from the end points where $C$ resp. $C^{-1}$ hits the
left resp. right vacuum. A more serious problem in the elliptic sector,
also observed in ref.\cite{LuS}, is that the  Liouville
exponentials
possess no natural restriction to the subspace of positive norm states,
given by the condition $|\varpi| <1+\pi/h , \ |\varpib|<1+\pi/h$.
For the coupling of $c<1$ matter to gravity, however, this problem
is irrelevant, as all negative norm states become
decoupled through the Virasoro conditions.

Let us  add here some remarks on the hyperbolic sector.
In this case, one should consider
the hermitian zero mode $P:=i\varpi$. However, the chiral vertex
operators shift $P$ formally by imaginary amounts. In order to have
a well-defined action of the zero mode shift operators $e^{-m\alpha_- q_0}$
in the hyperbolic sector, they should be applied to Gaussian wave packets
rather than momentum eigenstates\cite{BCGT}.
The former constitute a dense subset of the zero-mode
Hilbert space. Then for any Gauss packet $\langle q_0 | \psi_G \rangle=
N e^{-\beta (q_0-b)^2}$ with Re $\beta>0$, $e^{-m\alpha_- q_0} |\psi_G\rangle$
is again a Gauss packet, possessing a Fourier decomposition in terms of
real momenta $P$. Matrix elements in the zero mode space can thus be evaluated
by repeated Fourier transformation. Contour deformation considerations
 then show that
 in order  for some function $f(P)$ to fulfill
\beq
f(P)e^{-m\alpha_- q_0}=e^{-m\alpha_- q_0}f(P-2im)
\label{Pcommute}
\eeq
on the dense subspace, we need that $f(P)$ be analytic on the strip
Im $P \in [-2m,0]$ for $m>0$, resp. Im $P \in [0,-2m]$ for $m<0$,
with integrable singularities allowed on the real axis. Furthermore,
$f(P)$ needs to be exponentially bounded, $\lim_{|P| \to \infty}
\max_{-2m\le \hbox{Im} P \le 0} f(P)e^{-\beta P^2} =0  \ \forall \beta >0$
if $m>0$, and analogously if $m<0$. On the other hand, the coefficients
of Eq.\ref{Cbgas} contain poles at $P+i(2m-k)=0$ resp. $P+ik=0,
\ k=1,\dots J+m$. The commutation of these coefficients with shift operators
 would then generically produce unwanted
residue contributions in addition to the ``naive'' formula Eq.\ref{Pcommute},
and this would prevent us from directly taking over the results of the
locality analysis to the hyperbolic sector. It was observed in \cite{BCGT},
however, that the problem is absent for $J=1/2$ (provided $h<\pi$, which
is assumed anyway in the weak coupling sector).
One could think of defining the action
of shift operators resp. functions $f(P)$ through analytic continuation
from the elliptic sector; however, it seems that such attempts lead immediately
to problems with unitarity if the standard scalar product
$\langle P|P'\rangle
=\delta(P-P')$ is kept.
\subsection{The Liouville Field $\Phi$}
\subsubsection{Definition}
Having constructed Liouville exponentials with arbitrary continous spins,
we can now define the Liouville field $\Phi$ itself by\cite{OW}
\beq
\alpha_- \Phi := -{d\over dJ}\left.
e^{\textstyle -J\alpha_-\Phi}\right |_{J=0}
\label{4.1}
\eeq
Though $\Phi$ is not really a primary field --- it is similar to the
stress-energy tensor in this respect ---
it is needed to verify the validity of canonical commutation relations
and the quantum equations of motion\footnote{in their standard
form;
for a different approach, see ref.\cite{GS2}. }.
Thus we expand Eq.\ref{Cbgas} near $J=0$. In this limit, the factor
$\prod_{k=1}^{J+m}
\lfloor 1+2J-k\rfloor\to {2Jh\over \sin h}\prod_{k=2}^{J+m}
\lfloor 1+2J-k\rfloor$ vanishes  except for $J+m=0$, and  the exponential
tends to one as it should.
It then follows immediately that
$$
\Phi(\sigma, \tau)  =- (\vartheta_1(u) +\bar \vartheta_1(v) ) +
{2h\over \alpha_- \sin h }\times
$$
\beq
\sum_{n=1}^\infty  \tilde \mu_0^n {1 \over \lfloor n\rfloor}
\prod_{k=1}^n {1\over \lfloor \varpi+2n-k \rfloor \lfloor \varpi+k \rfloor}
S(u)^n \Sb(v)^n.
\label{4.2}
\eeq
\subsubsection{Periodicity properties and singularity structure}
In the hyperbolic sector where $\varpi=\varpib$, the Liouville field
of Eq.\ref{4.2} is manifestly periodic. However, inspecting the periodicity
behaviour of $\Phi$ in the elliptic sector with $\varpi \ne \varpib$, we find
that
\beq
\alpha_-\Phi(\sigma+2\pi,\tau)=\alpha_-\Phi(\sigma,\tau)-2\pi i k
\label{period}
\eeq
where $k$ is the parameter appearing in Eq.\ref{condx}. The constant is
entirely
produced by the free field contribution to $\Phi$, as the series in screening
charges is periodic order by order (cf. Eq.\ref{3.2}). Eq.\ref{period}
obviously calls for some explanation, as at least classically the Liouville
field should be periodic by definition. We will carry out the
discussion classically,
but this will suffice to obtain a qualitative understanding of the situation.
The essential point is that the definitions Eq.\ref{2.2} and Eq.\ref{4.1},
though seemingly equivalent classically, actually differ slightly in the
elliptic sector. If $2\sqrt\gamma \Phi(\sigma,\tau)$ is regular everywhere
in $[0,2\pi]$, the two definitions clearly can differ only by a constant. But
it is well known\cite{JKM} that in the elliptic sector (with $k\ne 0$) there
are
$|k|$ nonintersecting singularity lines, thus $|k|$ singularities in
$\sigma \in
[0,2\pi]$. At a singular point, the constant connecting the two definitions
may -  and does - change, creating in this way a nontrivial periodicity
behaviour of our field $\Phi$ of Eqs.\ref{4.1}, \ref{4.2}. Indeed, classically
the definition Eq.\ref{4.1} is equivalent to
\beq
2\sqrt\gamma\Phi=-2\ln {A'}^{-1/2}-2\ln B{B'}^{-1/2} -2\ln (1-A/B) +
\hbox{const}.
\label{classdefphi}
\eeq
with the series expansion representing the logarithm. It follows easily
from the results of \cite{JKM} that at each singular point, $2\ln (1-A/B)$
 - and hence $\Phi$ of Eq.\ref{classdefphi} - jumps by an imaginary
constant $-2\pi i \hbox{sgn} k$, whereas there is no such jump, of course,
in Eq.\ref{2.2} which is by definition real. Hence,
\[
2\sqrt\gamma(\Phi_{Eq.\ref{4.1}}(\sigma+2\pi,\tau) -\Phi_{Eq.\ref{4.1}}(\sigma,
\tau))
\]
\beq
-2\sqrt\gamma(\Phi_{Eq.\ref{2.2}}(\sigma+2\pi,\tau)-\Phi_{Eq.\ref{2.2}}
(\sigma,\tau)) =-2\pi ik
\label{phidiff}
\eeq
As the second difference is zero, we reproduce Eq.\ref{period}. Thus the
Liouville field we are using differs from the "true" one only by a constant
between any two singularity lines, but the constant changes at the
singularities. In particular, our $\Phi$ cannot be real everywhere in
$\sigma \in [0,2\pi]$. (This has nothing to do with the nonhermiticity
of the exponentials noted in section \ref{hermiticity}, as the latter is
independent of $\sigma$; indeed, for the exponentials with half-integer $J$,
the jumps play no role for the hermiticity behaviour). We remark that the
periodicity behaviour of our Liouville field is actually quite natural,
as the spectrum in the elliptic sector contains a winding number ($k$) and
therefore looks like that of a compactified field. Pursuing this analogy
further, we would be lead in the quantum case to impose the usual
single-valuedness condition on the field operators. In the free string case,
this enforces the quantization of the momenta, whereas here we obtain a
discretization of the spin,
\beq
Jk\in {\bf Z}.
\label{discretespin}
\eeq
Note that the periodicity behaviour of the (quantum) Liouville exponentials
can be read off directly from Eq.\ref{period}, though they are not actually
naive exponentials of $\Phi$. If we admit also anti-periodic behaviour of
the exponentials on the cylinder, then  we can  have
half-integer spins as well for $k$ odd.
One may speculate if the condition
Eq.\ref{discretespin} can be relaxed if we couple the theory to compactified
matter, in such a way that the multivaluedness of the Liouville part is
precisely
cancelled by that of the matter, but we will not go into this here.

When using the free field $\vartheta_2$ instead of $\vartheta_1$,
the periodicity behaviour of $\Phi$ of Eq.\ref{4.2} will be exactly opposite.
One may think that the regions of convergence for the two (classical)
expansions, $|A/B|<1$ resp. $|B/A|<1$ are complementary and therefore
there is no contradiction. However, in contrast to the hyperbolic sector we
have
$|A/B| \equiv 1$ in the elliptic sector, such that both series expansions are
exactly {\it on} their circle of convergence, and in fact converge there except
for the singularities at $A=B$. Thus we see that $\vartheta_1\leftrightarrow
\vartheta_2$ invariance is broken
by the singularities in the elliptic sector. However, for the exponentials
with half-integer $J$, we get the same periodicity behaviour for the
$\vartheta_1$
and the $\vartheta_2$ representation. Correspondingly, we were able even
in the quantum case to construct these exponentials in a $\vartheta_1
\leftrightarrow
\vartheta_2$ invariant way.
For continous $J$, however, it is not obvious how this
invariance can be
restored.

\subsubsection{The case $k=0$}
A special consideration is required for the case where $\varpi$ is real
(elliptic sector) but $k=0$ in Eq.\ref{condx}. It is known from the
work of \cite{JKM} that in this situation, there is no real Liouville field
even classically. However, from the point of view of the locality analysis,
the case $k=0$ is very natural and therefore we did not exclude it.
It is not hard to show that indeed more generally one needs to have
\beq
\varpi\varpib <0
\label{realitycond}
\eeq
classically in order to obtain real solutions of the Liouville equation with
positive cosmological constant. In the other case, one has $ 2\sqrt\gamma
\hbox{Im}\Phi =\pm i\pi$, and so the real part of $\Phi$ solves
 the Liouville equation
with negative cosmological constant. As regards the singularity structure,
$\sigma$ and $\tau$ essentially exchange their roles, and thus one obtains
 timelike instead of spacelike singularity lines\cite{P}.
The number of singularities can in general be greater than $|k|$, though
$|k|$ continues to characterize the periodicity behaviour of our solution
Eq.\ref{4.2}. The explanation is that the additional singularities always
come in pairs with opposite associated jumps $\pm 2\pi i$ of $2\sqrt\gamma
\Phi$, so their effect is not seen in the overall periodicity behaviour
of $\Phi$. In particular, for $k=0$ we now understand why the Liouville field
Eq.\ref{4.2} is periodic in spite of the possible presence of singularities.
We stress also that in our analysis there is no restriction on the sign of
$\varpi\varpib$, thus we describe solutions of the Liouville equation with
both signs of the cosmological constant.

\subsubsection{ The field equations}
For the considerations in this subsection and below, the similarity
transformations $T$ discussed above play no role and so we return to the
representation Eq.\ref{4.2} resp. Eq.\ref{Cbgas} of the Liouville exponentials.
Our first task is to compute $\partial_u \Phi$ and $\partial_v \Phi$.
For this we need an expression for
$\partial_u S^n\equiv \partial_u U_n^{(0)}$. Since
$U_n^{(0)}$ is a primary field with weight zero, its derivative is
primary with weight one. It is easy to see, by looking at the shift
properties in $\varpi$,  that it must be proportional to $U^{(-1)}_{n-1}$.
Thus we have
$$
\partial_u   U_n^{(0)}  =i c_n(\varpi) U^{(-1)}_{n}.
$$
In order to determine $c_n(\varpi) $,
one takes the matrix element between highest-weight states
and uses
 $\partial_u   U_n^{(0)} =i[ L_0,  U_n^{(0)}]$.  With the help of
Eq.\ref{3.6} for
 the normalization
factor $I_m^{(J)}$ and the expression for the Virasoro weights
$L_0 |\varpi> =((1+\pi/h)^2-\varpi^2) h/4\pi|\varpi>$, this gives
immediately
\beq
c_n(\varpi)=2\sin h q^{\varpi+1}\lfloor n\rfloor \lfloor \varpi+n\rfloor
\label{4.7}
\eeq
Thus we obtain the following formulae:
$$
\partial_u \Phi= -\partial_u \vartheta_1 +{2ih\over \alpha_- \sin h}
\sum_{n=1}^\infty
{\tilde \mu_0^{n} \over \lfloor n\rfloor}c_n(\varpi)\times
$$
\beq
\prod_{k=1}^{n}
{1 \over
\lfloor \varpi+2n-k \rfloor \lfloor \varpi+k \rfloor}
U^{(-1)}_{n} \Sb^n,
\label{duphi}
\eeq
$$
\partial_v \Phi= -\partial_v \varthetab_1 +{2ih\over \alpha_- \sin h}
\sum_{n=1}^\infty
{\tilde \mu_0^{n} \over \lfloor n\rfloor}\bar c_n(\varpi)\times
$$
\beq
\prod_{k=1}^{n}
{1 \over
 \lfloor\varpi+2n-k \rfloor \lfloor \varpi+k \rfloor}
S^n \Ub^{(-1)}_{n}.
\label{dvphi}
\eeq
In these two equations we observe the appearance of the coefficients of the
expansion Eq.\ref{Cbgas} with $J=-1$.  Taking the crossed derivative we
thus get
\beq
\partial_u \partial_v \Phi=-{\alpha_-\over 8}
e^{\textstyle \alpha_-\Phi} \ ,
\label{fequ}
\eeq
if we choose\hfill
\beq
\tilde \mu_0={1\over 32\pi \sin h}
\label{defmu0}
\eeq
Eq.\ref{fequ} is the quantum Liouville field equation associated
with an action given by Eq.\ref{2.1} with $2\sqrt{ \gamma}$ replaced by
$\alpha_-$, and  with $\mu=1$.
In view of the recently shown equivalence of different frameworks\cite{GS2},
we can directly compare this result with the one obtained in an
older  analysis
by Otto and Weigt\cite{OW}, and find agreement\footnote{
Note that the formula of Otto and Weigt quoted in ref.\cite{GS2} needs to be
multiplied by a factor $({\sin h \over h})^{2J}$ to be in accord with the
equations
of motion. (This was already noticed in \cite{OW}).}.
\subsubsection{Equal-time commutation relations}
We now proceed to the canonical commutation relations.
It is a trivial consequence of Eqs.\ref{loc},
\ref{4.1} that
\beq
[\Phi(\sigma,\tau),\Phi(\sigma',\tau)]=0.
\label{4.17}
\eeq
Next, by differentiating Eq.\ref{4.17} twice with respect to time
and using the equations of motion, we see that also
\beq
[\Pi(\sigma,\tau),\Pi(\sigma',\tau)]=0.
\label{4.18}
\eeq
where $\Pi(\sigma,\tau)$ is the canonical momentum,
\beq
\Pi(\sigma,\tau)={1\over 4\pi} \partial_\tau \Phi(\sigma,\tau)
\label{defPi}
\eeq
Furthermore we note that $[\Pi(\sigma,\tau), \Phi(\sigma',\tau)]$ can
be
nonvanishing only at $\sigma =\sigma'$, due to the fact the $R$-matrices
for arbitrary spins   depend on $\tau,\sigma$ only via the step functions
$\theta(u-u')$ resp. $\theta(v-v')$. On the other hand, the contribution
of the free field parts of $\Pi$ and $\Phi$ gives precisely the expected
result:
\beq
[\Pi(\sigma,\tau),\Phi(\sigma',\tau)]|_{\hbox{ free field}}
=-i\delta(\sigma-\sigma')
\label{4.19}
\eeq
We will show now that the sum of the other contributions vanishes.
First we observe that $[\Pi(\sigma,\tau),\Phi(\sigma',\tau)]=0$ order by
order in powers of screening charges, for $\sigma \ne \sigma'$ (the operators
$V_1^{(-1)}(u),{\overline V}_1^{(-1)}(v)$ appearing in $\Pi(\sigma,\tau)$ count
as screenings as well).
It suffices then to show that also at $\sigma=\sigma'$,
\beq
[\Pi(\sigma,\tau),\Phi(\sigma',\tau)]_N=0 \qquad \forall N\ge 1 \ ,
\label{orderN}
\eeq
where $[\Pi(\sigma,\tau),\Phi(\sigma',\tau)]_N$ denotes the $N$-screening
contribution
to the commutator.
Let us
represent the equal time commutator as a limit of time-ordered products,
\beq
[\Pi(\sigma,\tau),\Phi(\sigma',\tau)]_N=\lim_{\Delta \to 0}
\{\Pi(\sigma,\tau +\Delta)
\Phi(\sigma',\tau) -\Phi(\sigma',\tau)\Pi(\sigma,\tau-\Delta )\}_N
\label{4.20}
\eeq
as usual. Next we argue that for fixed $N$ and $h$ small enough,
the leading contributions to the operator
product appearing in Eq.\ref{4.20}
simply are not singular enough to give a contribution to the commutator
at $\sigma=\sigma'$. Indeed, we have\footnote{we use here the notation
$U_n^{(0)}$
instead of $S^n$ to make it clear that these operators exist for arbitrary $h$,
according to the remarks below Eq.\ref{3.6}.}
\[
\partial_u \vartheta_1(u') \cdot U_N^{(0)}(u) \sim N\alpha \ln(z-z')
U_N^{(0)}(u)
\]
with $z:=e^{iu}$, as well as
\beq
\vartheta_1(u) \cdot U_{N}^{(-1)}(u') \sim \alpha \ln(z-z')U_N^{(-1)}(u')
\label{4.21}
\eeq
Eq.\ref{4.21} is true even for arbitrary $h$. Finally, there are also terms
of the form
$U_n^{(0)}(u)U_{n'}^{(-1)}(u')$ , $n+n'=N$, which are finite
for $u \to u'$  if $h$ is small enough.
Similar contributions arise of course from the right-moving
parts. As the singularities are only logarithmic, they cannot produce
a nonvanishing distribution at $\sigma=\sigma'$.  Hence for fixed $N$
and small enough $h$, Eq.\ref{orderN} must be valid for all $\sigma,\sigma'$.
On the other hand, using the $R$-matrix Eq.\ref{3.20} and its right-moving
counterpart and ($0<\sigma<\sigma'<\pi, \  \tau=0$)
$$
\alpha_-[\vartheta_1(\sigma), V_1^{(-1)}(\sigma')] =-i\pi\alpha_-^2
V_1^{(-1)}(\sigma')
$$
$$
\alpha_-[\vartheta_1(\sigma),
S(\sigma')]=i\pi\alpha_-^2/( k(\varpi)-1) \,(2k(\varpi)S(\sigma)
+(1-3k(\varpi))S(\sigma'))
$$
\beq
\alpha_- [\partial_\tau \vartheta_1(\sigma),S(\sigma')]
=2\pi i \alpha_-^2
k(\varpi)V_1^{(-1)}(\sigma)
\label{thetacomm}
\eeq
it is easy to see that $[\Pi(\sigma,\tau),\Phi(\sigma',\tau)]_N$ for general
$h$
can be written in the form
\[
 [\Pi(\sigma,\tau),\Phi(\sigma',\tau)]_N=\sum_{l=0}^{N-1}\sum_{\bar l =0}^N
\beta_{l\bar l}(\varpi)U_l^{(0)}(u')U^{(-1)}_{N-l}(u) \Ub^{(0)}_{\bar l}(v')
 \Ub^{(0)}_{N-\bar l}(v)+
\]
\beq
\sum_{l=0}^{N}\sum_{\bar l =0}^{N-1}
\gamma_{l\bar l}(\varpi)U^{(0)}_{l}(u')U_{N-l}^{(0)}(u)
\Ub^{(0)}_{\bar l}(v') \Ub^{(-1)}_{N-1-\bar l}(v)
\label{commform}
\eeq
Clearly the coefficients $\beta_{l\bar l}$ and $\gamma_{l\bar l}$ are
analytic in $h$,
as the $R$-matrices and the coefficients on the r.h.s. of Eq.\ref{thetacomm}
have this property.
Since $\beta_{l\bar l}$ and $\gamma_{l\bar l}$ vanish for small enough $h$,
they
have to be zero everywhere. Thus we find for the full commutator,
\beq
[\Pi(\sigma,\tau),\Phi(\sigma',\tau)]=-i\delta(\sigma-\sigma')
\label{4.22}
\eeq
as expected, showing that the quantization scheme is indeed canonical.

\section{The case of two screening charges.  }
\subsection{The braiding}
In the above analysis, we have used only one quantum deformation
parameter $h$, which tends to zero in the classical limit $C\to \infty$
according to Eq.\ref{defh}. However, as explained in refs.\cite{CGR1}
one can consider also the operators $\Uhat_{\mhat}^{(\Jhat)}$ which
have the same form Eq.\ref{3.3} as the  $U_m^{(J)}$
but involve the deformation parameter
\beq
\hhat  \equiv \pi ^2/h
\label{2.35}
\eeq
Since $\Uhat_1^{(-1)}$ is also a dimension $1$ operator, the
$\Uhat _{\mhat}^{(\Jhat )}$ have a Coulomb gas representation completely
analogous to Eqs.\ref{2.1}-\ref{2.3}. Hence their exchange algebra (and
also their fusion properties) are the same as that of the ``unhatted''
operators. More generally, one can combine the fields
 $\Uhat_{\mhat}^{(\Jhat)}$ and $U_m^{(J)}$ by fusion
to obtain the operators $U_{m\mhat}^{(J\Jhat)}$
 with the properties (cf. Eqs.\ref{2.16}, \ref{3.4})
$$
\Delta_{J,\Jhat}=\Delta_{J+\Jhat\pi/h}
$$
\beq
U_{m\mhat}^{(J\Jhat)}\varpi =(\varpi +2m+2\mhat \pi/h)
U_{m\mhat}^{(J\Jhat)}\ ,\ U_{m\mhat}^{(J\Jhat)}\varpihat
=({\widehat \varpi} +2\mhat +2m h/\pi)
U_{m\mhat}^{(J\Jhat)}
\label{2.36}
\eeq
where $\widehat \varpi =\varpi  h/\pi$. Their normalization will be
discussed below. For half-integer $J,\Jhat$
the braiding (and fusion) of the $U_{m\mhat}^{(J\Jhat)}$
follows immediately from that of the $U_m^{(J)}$ because the two sets
of operators $U_m^{(J)}$ and $\Uhat_\mhat^{(\Jhat)}$ commute up to a phase:
\beq
U_m^{(J)}(\sigma) \Uhat_\mhat^{(\Jhat)}(\sigma')=e^{-2\pi iJ\Jhat
\epsilon(\sigma-\sigma')}\Uhat_\mhat^{(\Jhat)}(\sigma') U_m^{(J)}(\sigma)
\label{2.37}
\eeq
The natural expectation is that this will remain true even for
noninteger $2J$. However, here we meet a surprise.
 The commutation of $V_{-J}^{(J)}
(\sigma)$ and $\Vhat_{-\Jhat}^{(J)}(\sigma')$ gives the factor in
Eq.\ref{2.37}, the screening charges $S(\sigma)$ and $\hat S(\sigma')$
commute, but
\beq
S_{\sigma\sigma'} \Vhat_{-\Jhat}^{(\Jhat)}(\sigma')= e^{2\pi i\Jhat}
\Vhat_{-\Jhat}^{(\Jhat)}(\sigma')S_{\sigma\sigma'}
\label{2.38}
\eeq
whereas\hfill
\beq
S_\Delta \Vhat_{-\Jhat}^{(\Jhat)}(\sigma')= e^{-2\pi i\Jhat}\Vhat_{
-\Jhat}^{(\Jhat)}(\sigma')S_{\Delta }
\label{2.39}
\eeq
The phase factors agree only when $2J$ is integer, and thus the commutation
of hatted and unhatted operators becomes nontrivial in general.
We should therefore restart the machinery of section 2 with the
operators $U_{m\mhat}^{(J\Jhat)}$, where
\beq
U_{m \mhat}^{(J\Jhat )}:= V_{-J}^{(J)}\Vhat_{-\Jhat}^{(\Jhat)}
S^{J+m}{\hat S}^{\Jhat +\mhat}
\label{2.40}
\eeq
with the product of the first two factors being defined by renormalizing the
short-distance singularity  as usual (cf. also ref.\cite{GS1}).
Following exactly the same steps, we
arrive at the generalized version of Eq.\ref{2.16} (with shifted
$\varpi$ ):
\[
\prod_{s=1}^{J+m}(y'q^{2{\Je}'} +yq^{-(\varpi +2{\Je}'+2\Je+2s-1)})
\prod_{\hat s=1}^{\Jhat+\mhat}(\hat y'\qhat^{2{\Jehat}'} +\hat y
\qhat^{-(\varpihat +2{\Jehat}'+2\Jehat +2\hat s-1)})
\]
\[
\prod_{t=1}^{J'+m'} (y'+yq^{\varpi +2J+2m+2t-1})
\prod_{\hat t=1}^{\Jhat'+\mhat'} (\hat y'+\hat y
\qhat^{\varpihat +2\Jhat +2\mhat+2\hat t-1})=
\]
\[
\sum_{m_1\mhat_1} R(\underline{J},\underline{J}';\varpi+2\Je+2{\Je}')_{
\underline{m}_{\phantom {2}} \underline{m}'}^{\underline{m}_2
\underline{m}_1} q^{-\Je\Je'}\qhat^{-\Jehat{\Jehat}'}\times
\]
\[
\prod_{t=1}^{J'+m_2}(y'q^{2\Je} +yq^{\varpi +6\Je+2t-1})
\prod_{\hat t=1}^{\Jhat'+\mhat_2}(\hat y'\qhat^{2\Jehat}
+\hat y\qhat^{\varpihat +6\Jehat+2\hat t-1})
\]
\beq
\prod_{s=1}^{J+m_1} (y'+yq^{-(\varpi +2J'+2m_2 +2s -1)})
\prod_{\hat s=1}^{\Jhat+\mhat_1} (\hat y'+\hat y \qhat^{
-(\varpihat +2\Jhat'+2\mhat_2 +2\hat s -1)})
\label{2.41}
\eeq
where
\beq
\Je:=J+\Jhat \pi/h  \quad , \quad \Jehat := \Jhat + Jh/\pi
\label{2.42}
\eeq
At this point, it is useful to note that, since
$V_{-J}^{(J)}\Vhat_{-\Jhat}^{(\Jhat)}\propto V_{-\Je}^{(\Je)}$,
and since the above braiding
equations depend upon the $m$'s only through the screening numbers
$$
n=J+m,\quad n'=J'+m',\quad n_1=J+m_1, \quad n_2=J'+m_2,
$$
\beq
\nhat=\Jhat+\mhat,\quad \nhat'=\Jhat'+\mhat',\quad
\nhat_1=\Jhat+\mhat_1, \quad \nhat_2=\Jhat'+\mhat_2,
\label{ndef}
\eeq
 the $R$ matrix solution of Eq.\ref{2.41}
is only a function  of $\Je$ , $\Je'$, and of the screening numbers.
 In the continuous case
 $2J$ and $2\Jhat$ loose meaning, since it is not possible
to recover them
from $\Je$ (the same is of course also true for $\Je'$).
The screening numbers
 are such that $n+n'=n_1+n_2$, $\nhat+\nhat'=\nhat_1+\nhat_2$.
Returning to our main line, we see that we may factorize
the equation system into a product of one  system
of the type Eq.\ref{3.19} with another one of the same type
with $h\to \hhat$. This is realized if,  in Eq.\ref{3.19},   we
change  the  spins from  $J$ to $\Je$, and
replace the $m$'s by the quantities
$$
\ms= m-\Jhat {\pi\over h},\quad \ms'= m'-\Jhat' {\pi\over h},\quad
\msone= m_1-\Jhat {\pi\over h},\quad \mstwo= m_2-\Jhat' {\pi\over h}
$$
\beq
\mshat= \mhat-J {h\over \pi},\quad \mshat'= \mhat-J' {h\over \pi},\quad
\msonehat=\mhat_1-J {h\over \pi},\quad
\mstwohat= \mhat_2-J' {h\over \pi}.
\label{ms}
\eeq
This last replacement is such that the screening numbers
 remain unchanged,
that is $J+m=\Je+\ms$, and so on.
In  view of the previous
analysis, this tells us immediately what the solution of Eq.\ref{2.41} must be:
\beq
 R_U(\underline{J},\underline{J}';\varpi)_{
\underline{m}_{\phantom {2}} \underline{m}'}^{\underline{m}_2
\underline{m}_1}=q^{-\Je{\Je}'}\qhat^{-\Jehat{\Jehat}'}
R_U(\Je,{\Je}';\varpi )_{\ms {\ms}'}^{\mstwo \msone}
\hat R_U(\Jehat,{\Jehat}';\varpihat )_{\mshat {\mshat}'}
^{\mstwohat \msonehat}
\label{2.43}
\eeq
The $R$-matrices $R_U$ and $\hat R_U$
are given by the same expression
Eq.\ref{3.20}, written in terms of the deformation parameter $h$ resp. $\hhat$.
The phase factor is due to the fact that it is $q^{\Je{\Je}'}\qhat^{\Jehat
{\Jehat}'}\equiv q^{2\Je{\Je}'}$
rather than $q^{2\Je{\Je}'}\qhat^{2\Jehat{\Jehat}'}\equiv q^{4\Je{\Je}'}$
which appears in Eq.\ref{2.41}
(cf. Eq.\ref{3.19}). At this point, it is convenient to combine it with
the phase factor of the expression Eq.\ref{3.20} of $R_U$ and
the analogous one for $\hat R_U$. Altogether one gets
$$
q^{-2\Je {\Je}'} q^{{\ms}'(2\ms+{\ms}'+\varpi) -\mstwo(\mstwo + \varpi ) }
\qhat^{{\mshat}'(2\mshat + {\mshat}' +\varpihat)-
\mstwohat (\mstwohat+ \varpihat )}
$$
Using the fact that $J+m$, $\Jhat+\mhat$, etc. are integers,
this may may rewritten as
$$e^{ih\left [\varpi ({m'}^e-m_2^e)+
({m'}^e)^2 - (m_2^e)^2+2m^e{m'}^e\right ]}=
e^{-i\pi(\Delta_{x} +\Delta_{x+m^e_{1}+m^e_{2}}
-\Delta_{x+m^e} -\Delta_{x+m^e_2})}
$$
where we introduced
$$
m^e\equiv m+{\pi \over h} \mhat, \quad
{m'}^e \equiv m'+{\pi \over h} \mhat',
$$
\beq
m_1^e\equiv m_1+{\pi \over h} \mhat_1, \quad
m_2^e \equiv m_2+{\pi \over h} \mhat_2.
\label{me}
\eeq
This is precisely the factor expected from the general MS formalism, since
it involves the same linear combination of Virasoro  weights,
as say Eq.\ref{3.29}.
We can  write the complete $R$-matrix again in the
form  Eq.\ref{3.20}, if we introduce "vectorial"
quantities  $\kappa_{\Jgen1
,\Jgen2 }^{\Jgen{12} }$
and $\left\{ ^{\Jgen1 }_{\Jgen3 }\,^{\Jgen2 }_{\Jgen{123} } \right.
 \left |^{\Jgen{12} }_{\Jgen{23} }\right\}$ with $\Jgen{}  =(J,\Jhat)$ etc.
(cf. also ref.\cite{CGR1}).  Then
$$
R_U(\underline{J},\underline{J}';\varpi)_{
\underline{m}_{\phantom {2}} \underline{m}'}^{\underline{m}_2
\underline{m}_1}= e^{-i\pi(\Delta_{x} +\Delta_{x+m^e_{1}+m^e_{2}}
-\Delta_{x+m^e} -\Delta_{x+m^e_2})}
$$
$$
\times
{\kappa_{\Jgen{} ,\xgen{} +\mgen{12} }^{\xgen{} +\mgen2 }
\kappa_{\Jgen{} ',\xgen{} +\mgen2 }^{\xgen{} } \over
\kappa_{\Jgen{} ',\xgen{} +\mgen{12} }^{\xgen{} +\mgen{} }
\kappa_{\Jgen{} ,\xgen{} +\mgen{} }^{\xgen{} }}
\left\{ ^{\Jgen{} }_{\Jgen{} ' }\,^{\xgen +\mgen{12} }_{\xgen } \right.
 \left |^{\xgen +\mgen2 }_{\xgen +\mgen{} }\right\}
$$
where\hfill
\[
\kappa_{\Jgen{} ,\xgen +\mgen{} }^{\xgen }=\kappa_{\Je,x+\ms}^x
\kappahat_{\Jehat,\xhat +\mshat}^\xhat
\]
and\hfill
\beq
\left\{ ^{\Jgen{} }_{\Jgen{} ' }\,^{\xgen +\mgen{12} }_{\xgen }
\right.
 \left |^{\xgen +\mgen2 }_{\xgen +\mgen{} }\right\}=
\left \{^ {\Je }_{{\Je}' }\,^{x+\mssum}_{x} \right.
\left | ^{x+\mstwo }_{x +\ms }\right\}
\Gaghat\,   ^{\Jehat }_{{\Jehat}' }\, \,^{\xhat +\mssumhat }_{\xhat }
\bigr. \bverthat\, ^{\xhat +\mstwohat }_{\xhat +\mshat }  \Gadhat,
\label{2.44}
\eeq
where we have let $\mssum=\msone+\mstwo$, $\mssumhat=\msonehat+\mstwohat$.
Finally, we make contact with the form proposed
in ref.\cite{GR} by re-expressing the entries of the $6j$-symbols in terms
of the effective quantum numbers defined by Eq.\ref{me}. One easily finds,
using the double-brace notation of ref.\cite{GR}
$$
\left \{^ {\Je }_{{\Je}' }\,^{x+\mssum}_{x} \right.
\left | ^{x+\mstwo }_{x +\ms }\right\}
= \left \{^ {\Je }_{{\Je}' }
\,^{x+m^e_{12}-(\nhat_1+\nhat_2)\pi/h}_
{x\phantom{+m^e_{12}-(\nhat_1+\nhat_2)\pi/h}}
\right.
\left | ^{x+m^e_2-\nhat_2\pi/h }_{x +m^e-\nhat\pi/h }\right\}
=\left \{ \left \{^ {\Je }_{{\Je}' }\,
\,^{x+m^e_{12}}_
{x\phantom{+m^e_{12}}}
\right.
\left | ^{x+m^e_2 }_{x +m^e}\right\} \right\}
$$
\beq
\Gaghat\,   ^{\Jehat }_{{\Jehat}' }\, \,^{\xhat +\mssumhat }_{\xhat }
\bigr. \bverthat\, ^{\xhat +\mstwohat }_{\xhat +\mshat }  \Gadhat =
\Gaghat\,   ^{\Jehat }_{{\Jehat}' }\, \,
\,^{\xhat +\mhat^e_{12}-(n_1+n_2)h/\pi  }_
{\xhat\phantom{+\mhat^e_{12}-(n_1+n_2)h/\pi }}
\bigr. \bverthat\, ^{\xhat+\mhat^e_2-n_2 h/\pi  }
_{\xhat +\mhat^e-n h/\pi }  \Gadhat=
\Gaghat \Gaghat\,   ^{\Jehat }_{{\Jehat}' }\, \,
\,^{\xhat +\mhat^e_{12}  }_
{\xhat\phantom{+\mhat^e_{12} }}
\bigr. \bverthat\, ^{\xhat+\mhat^e_2  }
_{\xhat +\mhat^e }  \Gadhat \Gadhat
\label{2.44b}
\eeq
These  expressions coincide with the ones introduced in ref.\cite{GR}.
The complete $\kappa$ may also be written similarly as
\beq
\kappa_{\Jgen{} ,\xgen +\mgen{} }^{\xgen }=
\kappa_{\Je,x+m^e-(\Jhat+\mhat) \pi/h}^x\>
\kappahat_{\Jehat,\xhat +\mhat^e-(J+m) h/\pi}^\xhat
\label{kappa}
\eeq
Note that the hatted effective
quantum numbers are simply equal to the unhatted one multiplied by $h/\pi$.
Thus everything has been  expressed solely in terms of the effective
quantum numbers where hatted and unhatted quantum numbers cannot
be separated (clearly, the variable $x=(\varpi-\varpi_0)/2$ is also of
this type), but which are such  that the particular combinations
$\Je+\me$, $\Je'+\me'$, $\Je+m_1^e$, $\Je'+m_2^e$ may be written
as sums of a positive integer plus another positive integer times $\pi/h$.
This defines the screening numbers from the effective quantum numbers.
At this point one sees that the notation
$U_{m\mhat}^{(J\Jhat)}$ resp. $\Vt_{m\mhat}^{(J\Jhat)}$ used so
far is not
really appropriate for continuous spins, since $J$, $\Jhat$, $m$, $\mhat$
loose meaning. We will from now on indicate the true
parameters, that is write, e.g. $\Vt_{\Je+\me}^{(\Je)}$
(or equivalently $\Vt_{n\nhat }^{(\Je)}$, with $\Je+\me=n+\nhat \pi/h$)
 instead of
   $\Vt_{m\mhat}^{(J\Jhat)}$.
The outcome of the present discussion is that, if we define
$\Vt_{\Je+\me}^{(\Je)}$
 by letting
\beq
\Vt_{\Je+\me}^{(\Je)}:= \kappa_{\Jgen{} ,\xgen +\mgen{} }^{\xgen }
  U_{\Je+\me}^{(\Je)},
\label{2.45a}
\eeq
the braiding of the $\Vt$ fields is given by Eq.\ref{2.44} without the $\kappa$
factors, analogously to the case of a single screening charge, and
 thus coincides with the one written in
ref.\cite{GR}. This completes  the derivation of the
braiding of the chiral vertex operators with both
screening charges.

Finally let us check the orthogonality properties of the generalized
6-j symbols. Our aim is to prove that
\beq
\sum_{\Jgen{23} }
\left\{ ^{\Jgen{1} }_{\Jgen{3} }\,^{\Jgen{2} }_{\Jgen{123} }
\right.
 \left |^{\Jgen{12} }_{\Jgen{23} }\right\}
\left\{ ^{\Jgen{1} }_{\Jgen{3} }\,^{\Jgen{2} }_{\Jgen{123} }
\right.
 \left |^{\Kgen{12} }_{\Jgen{23} }\right\}=\delta_{\Jgen{12}, \Kgen{12} }.
\label{orthogene}
\eeq
What is the range of summation? In the  generalized 6-j symbols
$\left\{ ^{\Jgen{1} }_{\Jgen{3} }\,^{\Jgen{2} }_{\Jgen{123} }
\right.
 \left |^{\Jgen{12} }_{\Jgen{23} }\right\}$  there
are four pairs of screening numbers
\beqa
p_{1,2}+{\pi\over h} \phat_{1,2}&=& J^e_1+J^e_2-J^e_{12}, \quad
p_{12,3}+{\pi\over h} \phat_{12,3}= J^e_{12}+J^e_{3}-J^e_{123}\nnn
p_{2,3}+{\pi\over h} \phat_{2,3}&=& J^e_2+J^e_3-J^e_{23}, \quad
p_{1,23}+{\pi\over h} \phat_{1,23}= J^e_{1}+J^e_{23}-J^e_{123}
\label{pairs}
\eeqa
which are positive integers, such that $p_{1,2}+p_{12,3}=p_{2,3}+p_{1,23}$,
and $\phat_{1,2}+\phat_{12,3}=\phat_{2,3}+\phat_{1,23}$.
The two 6-j's  appearing
in Eq.\ref{orthogene} have the same $p_{2,3}$, $\phat_{2,3}$, $p_{1,23}$,
$\phat_{1,23}$. The other screening numbers are modified when
$\Jgen{12} $ is replaced by $\Kgen{12} $. We denote the result
 with the letter
$q$ instead of $p$. Since the screening number are linearly related, only
three of them are independent for each 6-j.
We may choose to eliminate the screening
numbers with index $12,3$. In the summation, only $J^e_{23}$ varies.
Equivalently, we can then sum over the positive integers $p_{2,3}$ and
$\phat_{2,3}$
since $J^e_{23}=J^e_2+J^e_3+J^e_{23}-p_{2,3}-(\pi/ h) \phat_{2,3}$.
    Let us next consider the l.h.s. of
Eq.\ref{orthogene} making use of the explicit expression Eq.\ref{2.44}.
It becomes
$$
\sum_{p_{2,3}\phat_{2,3}}
\left\{ ^{J^e_{1}\phantom{-(\phat_{2,3}+\phat_{1,23}){\pi\over h}  }}
_{J^e_{3}-(\phat_{2,3}+\phat_{1,23}){\pi\over h} }\,^{J^e_{2} }_{J^e_{123} }
\right.
 \left |^{J^e_{12}+\phat_{1,2}{\pi\over h} }
_{J^e_{23}-\phat_{1,23}{\pi\over h} }\right\}
\left\{ ^{J^e_{1}\phantom{-(\phat_{2,3}+\phat_{1,23}){\pi\over h}  }}
_{J^e_{3}-(\phat_{2,3}+\phat_{1,23}){\pi\over h} }\,^{J^e_{2} }_{J^e_{123} }
\right.
 \left |^{K^e_{12}+\qhat_{1,2}{\pi\over h} }
_{J^e_{23}-\phat_{1,23}{\pi\over h} }\right\}\times
$$
\beq
\Gaghat\,   ^{\Jhat^e_1\phantom{-(p_{2,3}+p_{1,23}){h\over \pi}   } }_
{\Jhat^e_2-(p_{2,3}+p_{1,23}){h\over \pi}}\, \,
\,^{\Jhat^e_2 }_
{\Jhat^e_{123}}
\bigr. \bverthat\, ^{\Jhat^e_{12}+p_{1,2}{h\over \pi} }
_{\Jhat^e_{23}-p_{1,23}{h\over \pi}}  \Gadhat
\Gaghat\,   ^{\Jhat^e_1\phantom{-(p_{2,3}+p_{1,23}){h\over \pi}  } }_
{\Jhat^e_2-(p_{2,3}+p_{1,23}){h\over \pi} }\, \,
\,^{\Jhat^e_2 }_
{\Jhat^e_{123}}
\bigr. \bverthat\, ^{\Khat^e_{12}+q_{1,2}{h\over \pi} }
_{\Jhat^e_{23}-p_{1,23}{h\over \pi}}  \Gadhat.
\label{6jshift}
\eeq
The shifts of the entries of the 6-j symbols are
not the same as in Eq.\ref{2.44b}. We used  the liberty of changing them
without changing the result ---  which was exhibited  in ref.\cite{GR} ---
to go to a more convenient expression. Now, we recall\cite{GR}  that these
shifts are precisely such that the screening numbers of the first
(resp. second) line only involve the $p$'s and $q$'s, (resp. the
$\phat$'s and $\qhat$'s),
so that the two sums  may be carried out independently.
Each  reduces  to an  orthogonality  relation of the type Eq.\ref{orth},
and this completes the derivation.
\subsection{The generalized Liouville field}
The generalization of the Liouville exponential  is given by
\beq
e^{\textstyle -J^e\alpha_-{\underline \Phi}(\sigma, \tau )}=
\sum _{n, \nhat =0}^{\infty} \mu_0^{n} \muhat_0^{\nhat}
\Vt^{(\Je)}_{n \nhat}(u)
{\overline \Vt^{(\Je)}_{n \nhat}}(v) \>
\label{genexp}
\eeq
where $\muhat_0$ is given by equations  similar to Eqs.\ref{mu0tilde}
and \ref{defmu0} with $h\to \hhat$. Of course, the preference of $\alpha_-$
over $\alpha_+$ in Eq.\ref{genexp} is purely notational as
$e^{-J^e\alpha_-{\underline \Phi}}
\equiv e^{-\Jhat^e \alpha_+{\underline \Phi}}$. Using
Eq.\ref{orthogene}, it is straightforward  to verify that the generalized
exponential  is local
and closed by fusion, provided $\varpi=\varpib$. The passage to $\vartheta_1
\leftrightarrow
\vartheta_2$ invariant exponentials (with $J,\Jhat$ positive half-integers)
proceeds
exactly as for the case of a single screening charge. The appropriate
similarity transformation $T$ can be read off directly from Eqs.\ref{fin},
\ref{C.5} of appendix \ref{connection}. It is equally straightforward
to extend the hermiticity discussion. We now turn to the generalized
Liouville field, which we
define again by Eq.\ref{4.1}. Thus we obtain
$$
{\underline \Phi}(\sigma, \tau)  =- (\vartheta_1(u) +\bar \vartheta_1(v) )
$$
$$
+{2h\over \alpha_- \sin h }
\sum_{n=1}^\infty  \tilde \mu_0^n {1 \over \lfloor n\rfloor}
\prod_{k=1}^n {1\over \lfloor \varpi+2 n-k \rfloor \lfloor \varpi+k \rfloor}
S(u)^n \Sb(v)^n.
$$
\beq
+{2\hhat\over \alpha_+ \sin \hhat }
\sum_{\nhat=1}^\infty  \tilde \muhat_0^n {1
\over \lfloorhat \nhat\rfloorhat}
\prod_{\hat k =1}^\nhat {1\over \lfloorhat \varpihat+2\nhat-\hat k \rfloorhat
 \lfloorhat \varpihat +\hat k \rfloorhat}
\Shat(u)^{\nhat} \Shatb(v)^{\nhat}.
\label{genLiou}
\eeq
It differs from the previous one by the last line. Since $\varpi=\varpib$,
$\Phi$ is periodic.
As a result, the
quantum field equation becomes
\beq
\partial_u \partial_v {\underline\Phi}=-{\alpha_-\over 8}
e^{\textstyle \alpha_-\Phi} -{\alpha_+\over 8}
e^{\textstyle \alpha_+\hat\Phi},
\label{genfequ}
\eeq
involving both  cosmological terms . Since ${\underline \Phi}=\Phi+\hat\Phi
+\vartheta_1+\bar \vartheta_1$, the validity of Eq.\ref{genfequ} is a trivial
consequence of the equations of motion with a single screening charge.
The reason why ${\underline \Phi}$ must be shifted w.r.t. $\Phi+\hat \Phi$ is
that
 $\Phi+\hat \Phi$ alone is not local. Indeed, using Eqs.\ref{kappashift} and
its
left-moving analog, and the fact that screening charges of different type
commute,
we see that the only nonzero contributions to
$[\Phi(\sigma,\tau)+\hat\Phi(\sigma,\tau),
\Phi(\sigma',\tau)+\hat\Phi(\sigma',\tau)]=[\Phi(\sigma,\tau),
\hat\Phi(\sigma',\tau)]+
[\hat\Phi(\sigma,\tau),\Phi(\sigma',\tau)]$ are of the form
$[\vartheta_1,\Shat]$ resp.
$[\vartheta_1,S]$. These commutators are precisely cancelled by the free field
shift.
Finally, one may easily extend the previous discussion of the canonical
commutation relations. The result is that Eqs.\ref{4.17}, \ref{4.18} and
\ref{4.22}
remain true  for the generalized Liouville field without any modification.

\section{Conclusions/Outlook}
The operator approach to Liouville Theory, which originated more than
ten years ago, has come a long way. Starting from the analysis of the
simplest Liouville field - the inverse square root of the metric - which
corresponds to the $J=1/2$ representation, it has now progressed
to the construction of the most general Liouville operators in the
standard (weak coupling) regime, corresponding to arbitrary
highest/lowest weight representations of the quantum group.
The underlying chiral algebra, either in its Bloch wave/Coulomb gas
or its quantum group covariant guise, has revealed beautiful structures
which may find applications also in very different contexts.
 Though the completion
of the quantization program of Gervais and Neveu thus finally comes into sight
for the weak coupling sector,  there remains an important complex of
questions yet to be addressed. While the Coulomb gas picture presented
here leads immediately to integral representations of arbitrary
n-point functions in the half-integer positive spin case\cite{LuS},
the correlators of operators with continous spins require more care.
This is due to the fact that outside the half-integer positive $J$ region,
the sums
representing the Liouville exponentials become infinite, and their evaluation
within correlation functions is quite nontrivial even in the simplest case
of the three-point function\cite{Cargese};  the study of the latter
 (for arbitrary
spins and central charge) is the subject of ongoing investigations.
Similarly, the hermiticity properties of the Liouville
operators and their invariance under the "large" $SL_2$ transformation
which exchanges $\vartheta_1$ and $\vartheta_2$ still pose an open
problem in the continous spin case.

In ref.\cite{G5}, where the computation
of three-point functions relevant for minimal matter coupled to gravity was
discussed (here $J$ is half-integer {\it negative}) , these difficulties were
avoided
in an interesting way by using instead of the "canonical" expression for the
Liouville exponential as given by Eq.\ref{genexp}
another operator with the same conformal
weight, which is however represented by a {\it finite} sum.
This approach appears
to be closer in spirit to the analytic continuation
procedure employed in the
path integral framework, and its connection
with the first-principle approach
along the lines
of this paper certainly deserves a better understanding. The Coulomb gas
picture for general spins should however open up the possibility of
studying arbitrary local (i.e. not integrated) correlation functions,
for any continous value
of the central charge, which is indispensable for a full understanding
of the integrability structure of Liouville theory as a local conformal
field theory in its own right. In contrast to this, the path integral or
matrix model approaches in their present form do not give any insight
into the local structure of the theory. In particular, the underlying
quantum group symmetry and its implications for the
structure of the operator algebra have so far remained completely
invisible in the other approaches.

The techniques used in this paper are applicable not only
to the standard weak coupling ($C>25$) regime considered here,
but also to the strong coupling theory as developed in refs.\cite{G3}
\cite{GR}. In this case, however, it turns out that one needs
inverse powers of the screening charges Eq.\ref{2.1}. These can be
immediately formulated in the Gervais-Neveu framework, as the
replacement $A\to -1/A$ which inverts the screening charge just
corresponds to exchanging the free field $\vartheta_1$ with $\vartheta_2$.
Unfortunately, the corresponding $V_m^{(J)}$ operators will then
again involve both free fields simultaneously, and the simple
commutation technique of section{3.1} cannot be used directly.
Nevertheless,  it should be possible to relate the braiding of
negative powers of the screenings to that of positive powers
making use of the known short-distance product of $S$ with its
inverse; we hope to address this question in a forthcoming publication.
\vskip 5mm
\noindent
{\bf Acknowledgements}

We are grateful to J.-F. Roussel for useful discussions. This work was
supported in part by the E.U. network "Capital Humain et Mobilit\'e",
contract \# CHRXCT920069.

\appendix
\section{ Some  previous results.}
\label{previous results}
The conventions are the same as in previous papers so that we will not
spell them out again.
In the operator approach, the quantum group structure was shown\cite{B,G1}
to be
of the type $U_q(sl(2))\odot U_\qhat (sl(2))$, where $h$ is given by
Eq.\ref{defh}, and
\beq
\hhat={\pi \over 12}\Bigl(C-13
+\sqrt {(C-25)(C-1)}\Bigr),
\label{2.17}
\eeq
Each quantum group parameter is associated with a screening charge by the
relations  $h=\pi (\alpha_-)^2/2$, $\hhat=\pi (\alpha_+)^2/2$.
The basic family of $(r,s)$ chiral operators in 2D gravity may be labelled
by two quantum group spins $J$ and $\Jhat$, with
$r=2\Jhat+1$, $s=2J+1$, so that
the spectrum of
Virasoro weights is given by
\beq
 \Delta_{J\Jhat}
={C-1\over 24}-
{ 1 \over 24} \left((J+\Jhat+1) \sqrt{C-1}
-(J-\Jhat) \sqrt{C-25} \right)^2,
\label{2.18}
\eeq
in agreement with Kac's formula.
One outcome of ref.\cite{CGR1} was the fusion and braiding of
the general chiral operators $V_{\mge}^{(\Jge )}$, also denoted
$V_{m \mhat}^{(J \Jhat)}$, where underlined  symbols denote double indices
$\Jge \equiv (J,  \, \Jhat)$, $\mge \equiv (m,\,  \mhat)$,  which were
all taken to be half integers:
$$
{\cal P}_{\Jgen{} } V_{\Jgen23 -\Jgen{} }
^{(\Jgen1 )}
V_{\Jgen3 -\Jgen23 }^{(\Jgen2 )} =
\sum_{\Jgen12 }
{g_{\Jgen1 \Jgen2 }^{\Jgen{12} }
\ g_{\Jgen{12} \Jgen3 }^{\Jgen{} }
\over
g _{\Jgen2 \Jgen3 }^{\Jgen{23} }
\ g_{{\Jgen1 }\Jgen{23} }^{\Jgen{} }
}
\left\{
^{\Jgen1 }_{\Jgen3 }\,^{\Jgen2 }_{\Jgen{} }
\right. \left |^{\Jgen{12} }_{\Jgen{23} }\right\} \times
$$
\beq
{\cal P}_{\Jgen{} } \sum_{\{\nu\}}
V_{\Jgen3 -\Jgen{} } ^{(\Jgen12 ,\{\nu\} )}
<\!\varpi _\Jgen12 ,\{\nu\}  \vert
V ^{(\Jgen1 )}_{\Jgen2 -\Jgen12 } \vert \varpi_{\Jgen2 }\! >.
\label{2.19}
\eeq
$$
{\cal P}_{\Jgen{} } V_{\Jgen23 -\Jgen{} }
^{(\Jgen1 )}
V_{\Jgen3 -\Jgen23 }^{(\Jgen2 )} =
\sum_{\Jgen13 }
e^{\pm i\pi (\Delta_\Jgen{} +\Delta_\Jgen3
-\Delta_{\Jgen23 }-\Delta_{\Jgen13 })}\times
$$
\beq
{g_{\Jgen1 \Jgen3 }^{\Jgen{13} }
\ g_{\Jgen{13} \Jgen3 }^{\Jgen{} }
\over
g _{\Jgen2 \Jgen3 }^{\Jgen{23} }
\ g_{{\Jgen1 }\Jgen{23} }^{\Jgen{} }
}
\left\{
^{\Jgen1 }_{\Jgen2 }\,^{\Jgen3 }_{\Jgen{} }
\right. \left |^{\Jgen{12} }_{\Jgen{23} }\right\}
{\cal P}_{\Jgen{} }
V_{\Jgen13 -\Jgen{} } ^{(\Jgen2 )}
V_{\Jgen3 -\Jgen13 }^{(\Jgen1 )},
\label{2.20}
\eeq
In these  formulae,  world-sheet variables are omitted,
and $\varpi$ is the rescaled zero-mode momentum of $\vartheta_1$ as in
Eq.\ref{defomega}. It characterizes the Verma modules
${\cal H}(\varpi)$, spanned by states noted $|\varpi, \, \{\nu\}>$,
where
$\{\nu\}$ is a multi-index. In the generic case, where the Verma
module is trivial, ${\cal H}(\varpi)$ is a Fock space generated by
the non-zero modes of the
free field $\vartheta_1$ (or equivalently of $\vartheta_2$),
with the  ground state
 $|\varpi>$.
  The symbol $\varpi_{\Jge}$ stands for
$\varpi_0+2J+2\Jhat \pi/h$ where $\varpi_0 =1+\pi/h$, and
 ${\cal P}_{\Jgen{} }$ is the projector on
${\cal H}(\varpi_{\Jge})$. The above formulae contain the recoupling
coefficients for the
quantum group structure $U_q(sl(2))\odot U_\qhat(sl(2))$, which
are defined by
\beq
\left\{
^{\Jgen1 }_{\Jgen3 }\,^{\Jgen2 }_{\Jgen{} }
\right. \left |^{\Jgen{12} }_{\Jgen{23} }\right\}
=
(-1)^{ \fusV 1,2,,23,3,12 }
\left\{
^{J_1}_{J_3}\,^{J_2}_{J}
\right. \left |^{J_{12}}_{J_{23}}\right\}
\gaghat
\,^{\Jhat_1}_{\Jhat_3}\,^{\Jhat_2}_{\Jhat}
\bigr. \bverthat \, ^{\Jhat_{12}}_{\Jhat_{23}}\gadhat
\label{2.22}
\eeq
where $\left\{
^{J_1}_{J_3}\,^{J_2}_{J}
\right. \left |^{J_{12}}_{J_{23}}\right\} $ is the $6j$ coefficient
associated with  $U_q(sl(2))$, while $\gaghat
\,^{\Jhat_1}_{\Jhat_3}\,^{\Jhat_2}_{\Jhat}
\bigr. \bverthat \, ^{\Jhat_{12}}_{\Jhat_{23}}\gadhat$ stands for the
$6j$  associated with $U_{\qhat}(sl(2))$. $\fusV 1,2,,23,3,12 $ is an
integer given by
\beq
\fusV 1,2,123,23,3,12
=2\Jhat_2(J_{12}+J_{23}-J_2-J_{123})
+2J_2(\Jhat_{12}+\Jhat_{23}-\Jhat_2-\Jhat_{123})
\label{sign}.
\eeq

In addition to these group
theoretic features  there appear the coupling constants
        $g_{{\Jgen{1} }\Jgen{2} }^{\Jgen{12} }$ whose expression
was given in ref.\cite{CGR1}. In order to connect with the
 general setting recalled in section \ref{self-contained},
let us indicate that $V^{(J, 0)}_{-J, 0}$ is proportional  to
$f_{-J}^{J}|_{\hbox {\scriptsize qu}}$,
and that $V^{(J, 0)}_{J, 0}$ corresponds to the normal
ordered exponential of $\vartheta_2$, that is to
$f_{J}^{J}|_{\hbox {\scriptsize qu}}$. One may
verify --- this is left as an exercise to the dedicated reader ---
that the equations just written are such that the braiding of these fields
is simply
\beqa
V^{(J_1, 0)}_{-J_1, 0}(\sigma_1)\>   V^{(J_2, 0)}_{-J_2, 0}(\sigma_2)&=&
e^{-2ih J_1 J_2\epsilon(\sigma_1-\sigma_2)}
 V^{(J_2, 0)}_{-J_2, 0}(\sigma_2)\>
 V^{(J_1, 0)}_{-J_1, 0}(\sigma_1), \label{2.23} \\
V^{(J_1, 0)}_{J_1, 0}(\sigma_1)\>  V^{(J_2, 0)}_{J_2, 0}(\sigma_2)&=&
e^{-2ih J_1 J_2\epsilon(\sigma_1-\sigma_2)}
V^{(J_2, 0)}_{J_2, 0}(\sigma_2)\>
 V^{(J_1, 0)}_{J_1, 0}(\sigma_1).
\label{2.24}
\eeqa
where $\epsilon(\sigma_1-\sigma_2)$ is the sign of $\sigma_1-\sigma_2$
(for definiteness, we consider the interval  $0\leq \sigma_i\leq \pi$).
This confirms that they are normal ordered exponentials  of
free fields, in agreement with the starting point of the
GN quantization.  Note that the braiding of $V^{(J, 0)}_{-J, 0}$ with
$V^{(J, 0)}_{J, 0}$ involves  the full complexity of the
6-j coefficients, so that the commutation relations of $\vartheta_1$
with $\vartheta_2$ are definitely not of   free-field type.
\section{Computation of the normalization integrals}
\label{normalization integrals}
\subsection{The case of a single screening charge}
To determine the normalization factors $I_m^{(J)}(\varpi)$ of Eq.\ref{3.6},
\beq
I_m^{(J)}(\varpi)=\langle \varpi |U_m^{(J)}(\sigma =0)|\varpi+2m\rangle
\label{A.1}
\eeq
we will reduce the integrals involved to those computed by Fateev and
Dotsenko\cite{DF}.
 Performing the Wick contractions in the standard
fashion, one obtains\cite{LuS}
$$
I_m^{(J)}(\varpi)=(-i)^n \oint dz_1 \cdots \oint dz_n
\prod_{j=1}^n (1-z_j)^\alpha z_j^\gamma
\prod_{j<k}(z_j-z_k)^\beta
$$
where\hfill
\beq
z_j=e^{i\sigma_j}, \ \alpha=2hJ/\pi, \ \beta =-2h/\pi, \ \gamma=(\varpi+2m-1)
h/\pi -1, \  n=J+m.
\label{A.2}
\eeq
The phase convention is that $(z_j-z_k)^\beta$ is real positive for
$0<z_k<z_j<1$, and otherwise defined by analytic continuation.
As usual,
this multiple integral is to be understood as the equal-time limit of a
time-ordered integral with $\sigma_j \to \sigma_j +i\tau_j$ and$\tau_j \to 0+$.
Thus the contours are given by the drawing
on the left of fig.1 .

\vspace{1cm}
\epsfig{file=gs3fig12.eps,width=12cm,clip=}
\centerline{fig1: Integration contours for Eq.A.2}
\vspace{1cm}

These contours are quite different from those appearing in the
Fateev-Dotsenko integrals\cite{DF}. To establish the connection, we
first compress the contours towards the real axis as shown on the
right of fig.1 ,
the distance between the contours and the real axis being infinitesimally
small. Then we split into contributions coming from the upper (UHP) resp.
lower half plane (LHP). The LHP parts differ from the UHP ones
only by
phase factors; hence $I_m^{(J)}$ can be reduced to a multiple integral
over the UHP contours only as in fig.2 .

\vspace{1cm}
\centerline{ \epsfig{file=gs3fig3.eps,width=6cm,clip=}}
\centerline{fig2: Equivalent set of integration contours}
\vspace{1cm}

Indeed, suppose the $z_n, \dots z_{n-p+1}$ integrations have already
been brought into the form of fig.2 . To continue with $z_{n-p}$, notice first
that $(z_j-z_{n-p})^\beta$ for $j<n-p$ takes the same values for $z_{n-p}$
infinitesimally above or below the real axis. Hence we need to
consider only
$(z_{n-p} -z_k)^\beta$, $k>n-p$. The permutation symmetry of the multiple
integral tell us in fact to consider together the set of $(p+1)!$
configurations
obtained by performing all permutations of a given fixed configuration
$z_n,\dots z_{n-p+1},z_{n-p}$. They can be organized into groups
$A_q^{(P)}$ and $B_q^{(P)}$ ($P$ labelling the permutation),
where $z_{n-p}$ occurs at the $q+1$ th resp. $p-q+1$ th position (cf. fig.3),

\vspace{1cm}
\epsfig{file=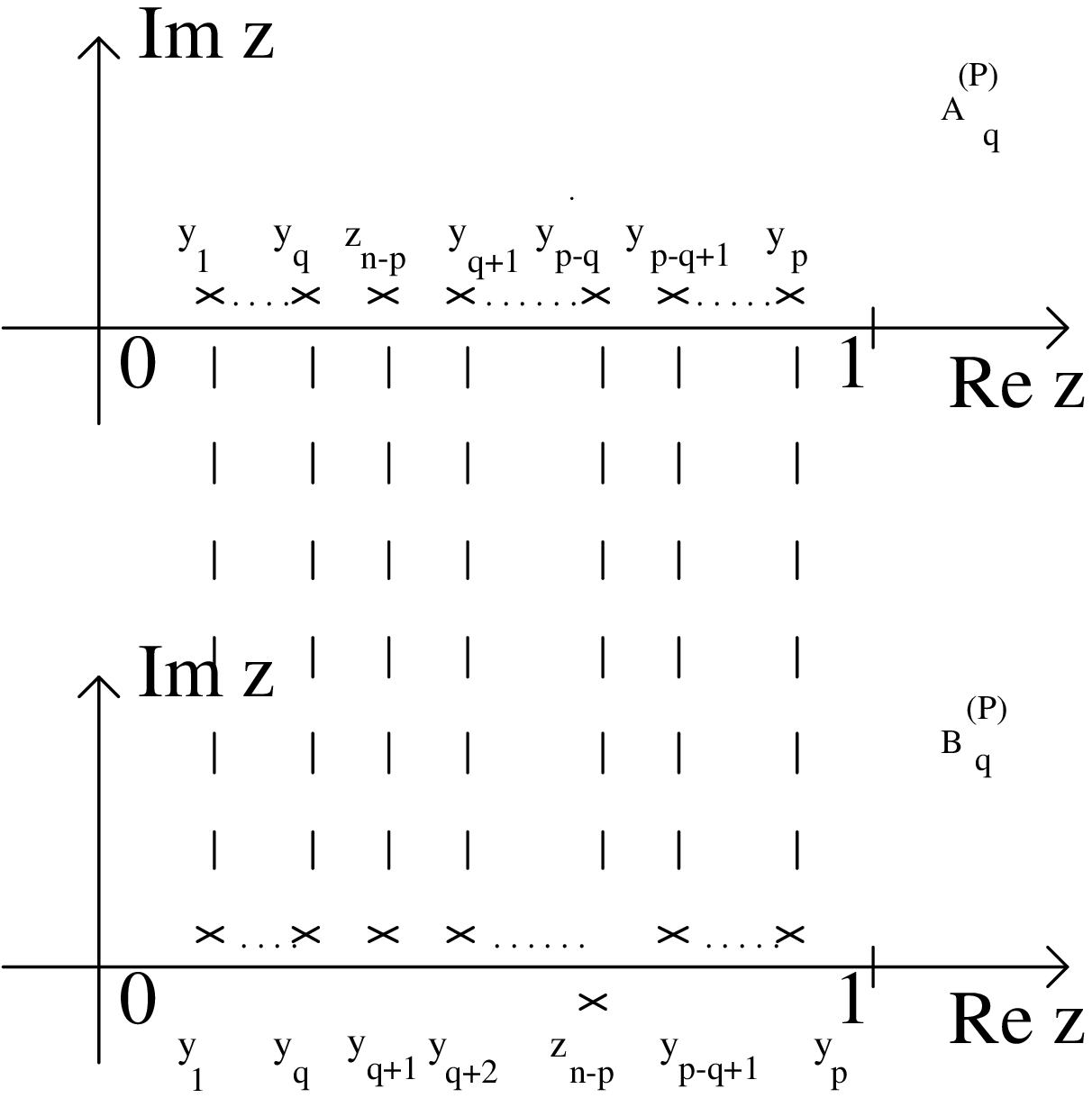,width=10cm,clip=}
\centerline{ fig3: Configurations of points to be compared for
$z_{n-p}$
on UHP resp. LHP}
\vspace{1cm}

with $y_1\cdots y_p=z_{P(n)},\dots z_{P(n-p+1)}$. Thus
$B_q^{(P)}$ differs from $A_q^{(P)}$ only by a permutation, except
that $z_{n-p}$ is now slightly below the real axis. It is now elementary
to add up the phases $\phi_{jk}$ of $(z_j-z_k), j<k$, for $A_q^{(P)}$
resp. $B_q^{(P)}$. If $j,k >n-p$ the phases generated by $A_q^{(P)}$
and $B_q^{(P)}$ agree since the relative ordering of the $y_i$
is the same.
For $j=n-p$, we have
\beq
\sum_{k>n-p}\phi_{n-p,k}|_{A_q^{(P)}}-\sum_{k>n-p}\phi_{n-p,k}
|_{B_q^{(P)}} =\pi(p-q)-(\pi p+2\pi(p-q)) =-\pi p,
\label{A.3}
\eeq
independent of $q$. Taking into account also the phase dependence of
$z_k^\gamma$, we find by successively applying the above argument for
$p=1,\dots n-1$:
\beq
I_m^{(J)}(\varpi)=(-i)^n\int_{C_1} dz_1 \dots \int_{C_n}dz_n
\prod_{j=1}^n (1-z_j)^\alpha z_j^\gamma \prod_{j<k} (z_j-z_k)^\beta
\prod_{l=0}^{n-1}(1-e^{2\pi i \gamma +i\pi l\beta})
\label{A.4}
\eeq
where the contours are as in fig.3 . Reversing the direction of integration and
renaming $z_1,\dots z_n \to z_n, \dots z_1$,
we obtain finally,
using $(z_k-z_j)^\beta =(z_j-z_k)^\beta e^{i\pi\beta}$,
$$
I_m^{(J)}(\varpi)=i^n\prod_{l=1}^n e^{i\pi\beta(l-1)} (1-e^{2\pi i\gamma
+i\pi\beta(l-1)}) J_{n0}(\alpha',\beta';\rho'),
$$
where\hfill
\beq
J_{n0}(\alpha',\beta';\rho')=\int_{\tilde C_1}dz_1 \dots \int_{\tilde C_n}
dz_n \prod_{j=1}^n (1-z_j)^\alpha z_j^\gamma \prod_{j<k} (z_j-z_k)^\beta
\label{A.5}
\eeq
$J_{n0}(\alpha',\beta';\rho')$ is precisely the integral appearing in
Eq.(A.4) of \cite{DF}, with $\alpha'=\gamma, \beta'=\alpha, 2\rho'=\beta$.
The contours $\tilde C_i$ are the same as in fig. 3 , except that all
directions are reversed, as well
as the order of $z_1, \dots z_n$ . Inserting the result of \cite{DF} for
$J_{n0}$,
one then obtains our formula Eq.\ref{3.6}.
\subsection{The relationship between $g_{J, x+m}^{x}$, $I_m^{(J)}$ and
$\kappa_{J, x+m}^{x}$}
In this part,  we verify the consistency of Eq.\ref{3.8}.
Let us compute $g_{J, x+m}^{x}/I_m^{(J)}(\varpi)$ from Eqs.\ref{defg},
\ref{3.8}.
It is convenient to combine the results in three factors.
$$
\prod_{k=1}^{n} {\sqrt{F[1+(2J-k+1)h/\pi]}\over \sqrt{F[1+kh/\pi]}}
{\Gamma [1+kh/\pi] \over \Gamma [1+(2J-k+1)h/\pi]}
=\prod_{k=1}^n {\sqrt{-\lfloor 2J-k+1\rfloor}\over \sqrt{ -\lfloor k \rfloor
}}$$
$$
\prod_{k=1}^{n} \sqrt{ F[(\varpi+2m-k)h/\pi]}
\Gamma [1-(\varpi+2m-k)h/\pi]=
{\pi^{n/2}\over \prod_{k=1}^{n} \sqrt{ \sin [ h (\varpi+2m-k))]}}
$$
\beq
\prod_{k=1}^{n} \sqrt{ F[-(\varpi+ k)h/\pi]}
\Gamma [1+(\varpi+k)h/\pi]=
{\pi^{n/2}\over \prod_{k=1}^{n} \sqrt{ -\sin [ h (\varpi+k))]}}
\label{consist}
\eeq
These equations easily lead to Eq.\ref{3.9}\footnote{In order to define
the square roots in the first of Eqs.\ref{consist}, we give a small
negative imaginary part to h (cf. ref.\cite{G5}). Then we can actually
cancel the minus
signs on the r.h.s. of this equation.}.

\subsection{The case of two screening charges}
Up to now we have been considering only the case of a single screening
charge. However, the Fateev-Dotsenko formulae also cover  the general
case with both screening charges, corresponding to the normalizations
$I_{m\mhat}^{(J\Jhat)}(\varpi)$. The above calculation is easily
generalized and we skip the details. The final result can be written
\[
I_{m\mhat}^{(J\Jhat)}(\varpi)=I^{(\Je)}_{\ms}(\varpi+2\nhat\pi/h)
{\widehat I}^{(\Jehat)}_{\mshat}
(\varpihat +2nh/\pi) (i\pi/h)^{2n\nhat}\times
\]
\[
\prod_{l,\hat l =1}^{n,\nhat}\{ (l+\hat l \pi/h)(\varpi +2m +2\mhat\pi/h
-l-\hat l\pi/h)(\varpi+l+\hat l \pi/h)
(2\Je -(l-1)-(\hat l -1)\pi/h)\}^{-1}
\]
\beq
\times\prod_{\hat l=1}^{2\nhat}\prod_{l=1}^n (\hat l +(\varpi +l)h/\pi)
\prod_{ l=1}^{2n}\prod_{\hat l=1}^\nhat ( l +(\varpihat +\hat l)\pi /h)
\label{A.6}
\eeq

\section{The earlier definition of the Liouville exponential}
\label{connection}
For completeness, let us connect Eq.\ref{3.31} with the previous definition
of ref.\cite{G5} for the half-integer case. In this article the
following formula was introduced (we distinguish it by an index G.)
\begin{equation}
\left. e^{\textstyle -J\alpha_-\Phi(\sigma, \tau )}\right |_{\hbox{G.}}=
\tilde c_J {1\over \sqrt{\lfloor  \varpi \rfloor}}\>\sum _{m=-J}^J
\,
 C_m^{(J)}(\varpi)
\psi_m^{(J)}(z)\,
{\overline \psi_{m}^{(J)}}(\zb) \sqrt{\lfloor  \varpi \rfloor},
\label{B.1}
\end{equation}
where $\tilde c_J$ is an arbitrary constant, and
\beq
C_m^{(J)}(\varpi)=(-1)^{J-m}\, (2i \sin h )^{2J} e^{ihJ} {2J
\choose J-m}\,
{\lfloor \varpi-J+m \rfloor_{2J+1}\over \lfloor \varpi+2m
\rfloor}.
\label{B.2}
\eeq
 The fields $\psi_{m}^{(J)}$ are similar to the
$\Vt_{m}^{(J)}$'s, but have a different normalization:
$$
\psi_{m}^{(J)}=E_{m}^{(J)} V_{m}^{(J)},
$$
with\hfill
$$
E_m^{(J)}(\varpi)={\prod_{r=1}^{2J}\Gamma(1+rh/\pi)\over
\prod_{r=1}^{J-m}\Gamma(1+rh/\pi)\prod_{r=1}^{J+m}\Gamma(1+rh/\pi)}
\prod_{r=1}^{2m}{\sqrt{\Gamma[(\varpi+r-1)h/\pi]}\over
\sqrt{\Gamma[-(\varpi+r)h/\pi]}} \times
$$
\beq
\prod_{r=1}^{J-m}\Gamma[(\varpi-r)h/\pi] \prod_{r=1}^{J+m}
\Gamma[-(\varpi+r)h/\pi]
\label{emj}
\eeq
For $\varpi=\varpib$, the case considered in ref.\cite{G5}, elementary
manipulations lead to the formula
\beq
C_m^{(J)}(\varpi){E_m^{(J)}(\varpi)\bar E_m^{(J)} (\varpi) \over
g^x_{J,x+m}g^x_{J,x+m}}=(2i\pi^2/h)^{2J}e^{ihJ} {\Gamma(\varpi+2m+1)
\Gamma[(\varpi+2m)h/\pi]\over
\Gamma(\varpi+1)\Gamma(\varpi h/\pi)}
\label{C.3}
\eeq
Thus we can write $\left. e^{\textstyle -J\alpha_-\Phi(\sigma, \tau )}
\right |_{\hbox{G.}}$
in the form
$$
\left. e^{\textstyle -J\alpha_-\Phi(\sigma, \tau )}\right |_{\hbox{G.}}=
c_J{\rho^2(\varpi)
\over \sqrt{\lfloor \varpi\rfloor}}\>\sum_{m=-J}^J \, \Vt_m^{(J)}\Vtb_m^{(J)}
{\sqrt{\lfloor\varpi\rfloor}\over \rho^2(\varpi)}
$$
with\hfill
\beq
\rho^2(\varpi)={\sqrt{h/\pi} \over \Gamma(\varpi h/\pi) \Gamma(\varpi+1)}=
{1\over \Gamma(\varpi h/\pi) \Gamma(\varpihat \pi/h) \sqrt{\varpi\varpihat}}
\label{C.4}
\eeq
Thus the previous definition is related to Eq.\ref{3.31}
 by a transformation
of the type we have discussed in detail in ref\cite{GS2}, that is a
transformation of the Hilbert space that only involves the zero-mode
$\varpi$. We see that when written in the form Eq.\ref{C.4}, the previous
definition
Eq.\ref{B.1} has meaning even for arbitrary $J$. This was already noticed in
ref.\cite{GS2}, where the relation with other approaches\cite{BCGT}\cite{OW}
was established by transformations of the same type.

Finally, let us deal with the case of two screening charges. In the
half-integer case, the most general Liouville field was defined by
\beq
\left. e^{\textstyle -(J\alpha_-+\Jhat\alpha_+) \Phi}\right |_{\hbox{G.}}
\sim
\left. e^{\textstyle -J\alpha_-\Phi}\right |_{\hbox{G.}}
\left. e^{\textstyle -\Jhat \alpha_+\Phi}\right |_{\hbox{G.}}
\label{fin}
\eeq
where $\sim$ means that one keeps  the leading-order term in the fusion.
According to ref.\cite{CGR1}, the leading order fusion of the $\Vt$ fields
is simply, for half-integer spins,
 $\Vt_{m}^{(J)} \Vthat_{\mhat}^{(\Jhat)}\sim \Vt_{m \mhat}^{(J \Jhat)}$.
Thus we get after commuting the intermediate  $\sqrt{\lfloor\varpi\rfloor}$
resp. $\sqrt{\lfloorhat \varpihat\rfloorhat}$ factors to the right resp.
to the left and redefining the normalization constant,
$$
\left. e^{\textstyle -(J\alpha_-+\Jhat\alpha_+) \Phi}\right |_{\hbox{G.}}=
c_\Jge {\rho(\varpi)^2 \over \sqrt{\lfloor  \varpi \rfloor}
\sqrt{\lfloorhat \varpihat\rfloorhat}}\times
$$
\beq
\sum _{m, \mhat } (-1)^{J+m} (-1)^{\Jhat+\mhat}
\Vt_{m \mhat}^{(J \Jhat)}(u)\,
{\overline \Vt}_{m \mhat}^{(J \Jhat)}(v)
{\sqrt{\lfloor  \varpi \rfloor}
\sqrt{\lfloorhat \varpihat\rfloorhat}\over
\rho(\varpi)^2},
\label{C.5}
\eeq
which is related to Eq.\ref{genexp} in a way similar to the
case of a single screening charge. In contrast to Eq.\ref{C.4}, we here
have $\mu_0=\muhat_0=-1$, but it is also possible to have $\mu_0=\muhat_0=1$
if we drop the $\sqrt{\lfloor \varpi\rfloor}$ factors in the definition
Eq.\ref{B.1} of the old Gervais exponentials, and its hatted counterpart.
Note that it was crucial to define
$\rho(\varpi)$ to be symmetric between $h$ and $\hhat$, so that the
$\rho$ factors
 in between the two exponentials  could cancel in Eq.\ref{fin}.  For general
continous spins, $\Vt^{(J^e)}_{n\nhat}$ and
$\Vt_m^{(J)} \Vthat_\mhat^{(\Jhat)}$
differ by a nontrivial normalization factor, and Eq.\ref{genexp} must be used
instead of Eq.\ref{fin}.


\begin{thebibliography}{99}
\bibitem{B} O. Babelon,
\pl   B215, 1988, 523 .
\bibitem{G1} J.-L. Gervais,  \cmp 130, 1990, 257 .
\bibitem{G2} J.-L. Gervais, \pl B243, 1990, 85 .
\bibitem{CG2} E. Cremmer, J.-L. Gervais,
 \cmp 144, 1992, 279 .
\bibitem{G4} J.-L. Gervais, \ijmp 6, 1991, 2805 .
\bibitem{G3} J.-L. Gervais, \cmp 138, 1991, 301 .
\bibitem{G5} J.-L. Gervais, ``Quantum group derivation
of 2D gravity-matter coupling'' Invited talk at
the Stony Brook meeting {\sl String and Symmetry 1991},
LPTENS preprint 91/22,\np B391, 1993, 287 .
\bibitem{CGR1} E. Cremmer, J.-L. Gervais,
J.-F. Roussel, \np B413, 1994, 244.

\bibitem{CGR2} E. Cremmer, J.-L. Gervais,
J.-F. Roussel, \cmp 161, 1994, 597.

\bibitem{GN4} J.-L. Gervais, A. Neveu, \np B238, 1984, 125;
ibid., p. 396.
\bibitem {GS1} J.-L. Gervais, J. Schnittger,
\pl B315, 1993, 258 \hfill

\bibitem{MS} G. Moore, N. Seiberg, \cmp 123, 1989, 77.

\bibitem{GR} J.-L. Gervais, J.-F. Roussel, ``Solving the
strongly coupled 2D gravity: 2. fractional spin operators,
and topological three-point functions'' preprint LPTENS 94/01,
hep-th/9403026
\bibitem{ANPS} A. Anderson,
B.E.W. Nilsson, C.N. Pope and K.S.Stelle,
``The Multivalued Free-field Maps of Liouville and Toda Gravities''
 preprint CTP TAMU-1/94, Goteborg ITP-94-3, Imperial/TP/93-94/13,
hep-th/9401007.
\bibitem{GN3} J.-L. Gervais, A. Neveu,  \np B224, 1983, 329.
\bibitem{GN2} J.-L. Gervais, A. Neveu,  \np B209, 1982, 125.
\bibitem {LuS} D. L\"ust, J. Schnittger,
\ijmp A6, 1991, 3625; J. Schnittger,  Ph.D. thesis,
 Munich 1990.\hfill
\bibitem{JKM} L. Johannson, A. Kihlberg, R. Marnelius,
\prd 29, 1984, 2798;
L. Johannson, R. Marnelius,
\np B254, 1985, 201.\hfill
\bibitem{GS2} J.-L. Gervais, J. Schnittger,
\np B413,  1994, 277.
\bibitem{OW} H.J. Otto, G. Weigt,
\pl B159, 1985, 341; {\sl Z. Phys. } {\bf C31},
(1986) 219; G. Weigt, ``Critical exponents of conformal fields
coupled to two-dimensional quantum gravity in the conformal gauge'',
talk given at 1989 Karpacz Winter School of Theor. Physics, preprint
PHE-90-15; G. Weigt, ``Canonical quantization of the Liouville theory,
quantum group structures, and correlation functions'', talk given at 1992
Johns Hopkins Workshop on Current Problems in Particle Theory, Goteborg,
Sweden, hep-th 9208075, print-92-0383 (DESY-IFH).\hfill
\bibitem{BCGT} E.Braaten, T.Curtright, C.Thorn,
\pl B118, 1982, 115;
\prl  48, 1982, 1309;
\annp 147, 1983, 365;
 E.Braaten, T.Curtright, G.Ghandour, C.Thorn,
\prl 51, 1983, 19;
\annp 153, 1984, 147.\hfill
\bibitem{P} G. Jorjadze, A. Pogrebkov, M. Polivanov,
{\sl Teor. Mat. Fiz.}  {\bf 40} (1979) 221;
A. Pogrebkov, {\sl Teor. Mat. Fiz.} {\bf 45} (1980) 161;
G. Jorjadze, A. Pogrebkov, M. Polivanov, S. Talalov,
{\sl J. Phys. A: Math. Gen.}  {\bf 19} (1986) 121.
\bibitem{BP} J. Balog and L. Palla,
\pl B274, 1992, 232.
\bibitem{Cargese} J. Schnittger, proceedings of the
1993 NATO adv. research workshop
on new developments in string theory,
conformal models and
field theory, Cargese May 93,
Plenum Press, to appear.\hfill
\bibitem{DF}  Vl. Dotsenko, V. Fateev,
\np B251, 1985, 691 .
\bibitem{Fe} G. Felder, \np B317, 1989, 215; erratum \np B324, 1989, 548.




\end{thebibliography}
\end{document}